\documentclass[12pt]{article}
\usepackage{a4wide}
\usepackage{latexsym}
\usepackage{amsmath}
\usepackage{amsfonts}
\usepackage{amscd}
\usepackage{cite}
\usepackage{placeins}

\usepackage{pslatex}
\usepackage{graphicx}
\usepackage[latin1,utf8]{inputenc}
\usepackage[T1]{fontenc}

\allowdisplaybreaks

\newcommand{\bq}{\begin{eqnarray}}
\newcommand{\eq}{\end{eqnarray}}
\newcommand{\eps}{\varepsilon}
\newcommand{\Nletter}{N_{\mathrm{letter}}}

\begin{document}

\thispagestyle{empty}

\begin{flushright}
  MITP/22-073
\end{flushright}

\vspace{1.5cm}

\begin{center}
  {\Large\bf Three-loop master integrals for the Higgs boson self-energy with internal top-quarks and W-bosons \\
  }
  \vspace{1cm}
  {\large Ekta Chaubey ${}^{a}$, Ina H\"onemann ${}^{b}$ and Stefan Weinzierl ${}^{b}$ \\
  \vspace{1cm}
      {\small \em ${}^{a}$ Physics Department, University of Turin and INFN Turin, }\\
      {\small \em Via Pietro Giuria 1, I-10125, Turin, Italy}\\
 \vspace{2mm}
      {\small \em ${}^{b}$ PRISMA Cluster of Excellence, Institut f{\"u}r Physik, }\\
      {\small \em Johannes Gutenberg-Universit{\"a}t Mainz,}\\
      {\small \em D - 55099 Mainz, Germany}\\
  } 
\end{center}

\vspace{2cm}

\begin{abstract}\noindent
  {
We consider the full set of master integrals with internal top-and $W$-propagators contributing to the 
three-loop Higgs self-energy diagrams of order ${\mathcal O}(\alpha^2 \alpha_s)$.
We split the master integrals into a system relevant to the Feynman diagrams
proportional to the product of Yukawa couplings $y_b y_t$
and the complement.
For both systems we define master integrals of uniform weight, such that the associated differential equation is 
in $\varepsilon$-factorised form.
The occurring square roots are rationalised and
all master integrals are expressible in multiple polylogarithms.
   }
\end{abstract}

\vspace*{\fill}

\newpage

\section{Introduction}
\label{sect:intro}

Higgs precision physics is at the core of the experimental LHC program. 
On the theoretical side this requires higher-order calculations in 
perturbation theory \cite{Lepage:2014fla,Anastasiou:2015vya,Bonciani:2016qxi,Mistlberger:2018etf,Kara:2018fwh,Duhr:2019kwi,Ajjath:2019ixh,Aparisi:2021tym,Martin:2022qiv,Spira:2019iec,Chakraborty:2022yan,Kataev:2022dua}.
The Higgs boson self-energy is an essential building block for many observables.
Through renormalisation it enters the Higgs mass definition.
Furthermore, the imaginary part of the Higgs boson self-energy is directly related through the optical theorem to the 
Higgs decay rate \cite{Kataev:1993be,Chetyrkin:1995pd,Chetyrkin:1997mb,Chetyrkin:1997vj,Baikov:2005rw,Bernreuther:2018ynm,Primo:2018zby,Kataev:1997cq,Kwiatkowski:1994cu,Kniehl:1994ju,Mihaila:2015lwa,Butenschoen:2006ns,Butenschoen:2007hz,Anastasiou:2011qx,Mondini:2019gid,Mondini:2020uyy,Behring:2019oci}.

In this article we consider three-loop contributions with internal top-and $W$-propagators to the Higgs boson self-energy.
Due to the presence of the internal top- and $W$-propagators 
the corresponding Feynman diagrams are among the more complicated Feynman integrals contributing at this order in perturbation theory.
These integrals have been computed for $p^2=m_H^2$ approximately as an expansion in $m_H^2/m_t^2$
in ref.~ \cite{Mihaila:2015lwa}.

In this paper we present the analytic calculation of the relevant master integrals.
We keep the full dependence on the external momentum squared $p^2$ and 
the heavy particle masses ($m_t$ and $m_W$), but neglect the $b$-quark mass.
The Feynman integrals depend on two dimensionless variables, which we may take initially as
\bq
 v \; = \; \frac{p^2}{m_t^2},
 & &
 w \; = \; \frac{m_W^2}{m_t^2}.
\eq
It is convenient to divide the Feynman integrals into two sets:
A set of $105$ master integrals corresponding to the set of Feynman diagrams
being proportional to the product of Yukawa couplings $y_b y_t$
and the complement, consisting of $33$ additional master integrals.
We treat the first set of $105$ master integrals in the main part of the paper
and the $33$ additional integrals in appendix~\ref{sect:additional_topologies}.
The reason is that the sets of occurring square roots differ. 
In total we encounter four different square roots $r_1$-$r_4$.
While we are not able to rationalise simultaneously all four square roots, we may set up
three independent systems of differential equations, such that within one system all square roots
can be rationalised.

In the main part of the paper we treat the set of $105$ master integrals corresponding to the set of Feynman diagrams
being proportional to the product of Yukawa couplings $y_b y_t$.
We define a basis of master integrals of uniform weight.
In this basis, the associated differential equation is $\eps$-factorised and all occurring differential one-forms 
are dlog-forms.
In this system we encounter two square roots, which can be rationalised simultaneously.
This is done by a change of variables from $(v,w)$ to $(x,y)$.
The variables $x$ and $y$ will be defined in the main part of this paper.
This change of variables allows us to express all master integrals to any order in the dimensional regularisation parameter $\eps$
in terms of multiple polylogarithms.
The alphabet for the multiple polylogarithms consists of $21$ letters.

The remaining master integrals, relevant to Feynman diagrams with internal top- and $W$-propagators, where
the Higgs boson couples either to a $W$-boson and a $b$-quark or to two $b$-quarks
are discussed in appendix~\ref{sect:additional_topologies}.
The calculation proceeds exactly along the same lines as for the first set.
The only difference is that additional square roots occur, which require different rationalisations.

This paper is organised as follows:
In section~\ref{sect:notation} we introduce our notation for the three-loop integrals.
The master integrals of uniform weight contributing to Feynman diagrams proportional to the Yukawa couplings $y_b y_t$
are presented in section~\ref{sect:masters}.
The differential equation for these master integrals is in $\eps$-factorised form.
The differential equation is discussed in section~\ref{sect:differential_forms}.
The solution of the differential equation requires boundary values, which we give in
section~\ref{sect:boundary}.
Analytical results for the master integrals are presented in section~\ref{sect:analytical_results}.
Numerical results are presented in section~\ref{sect:numerical_results}.
Finally, our conclusions are given in section~\ref{sect:conclusions}.
Appendix~\ref{sect:additional_topologies} presents the additional master integrals needed for the contributions not proportional to the Yukawa couplings $y_b y_t$.
In appendix~\ref{sect:supplement} we describe the content of the supplementary electronic file
attached to the arxiv version of this article.


\section{Notation}
\label{sect:notation}

\begin{figure}
\begin{center}
\includegraphics[scale=1.0]{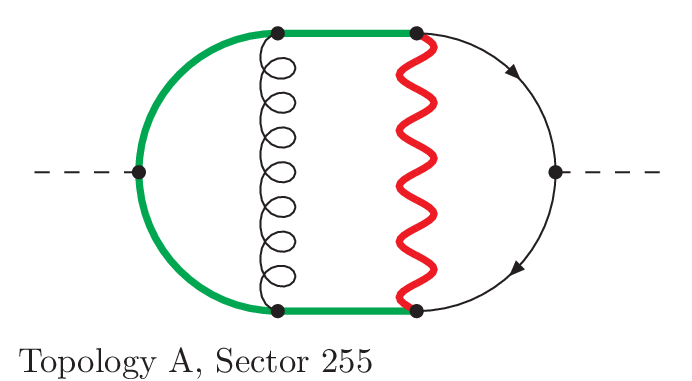}
\hspace*{10mm}
\includegraphics[scale=1.0]{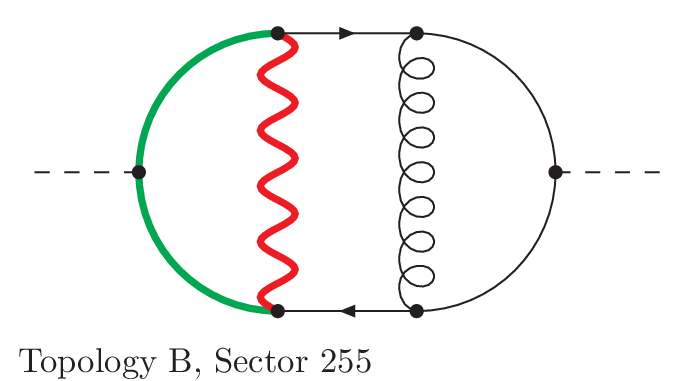}
\\
\includegraphics[scale=1.0]{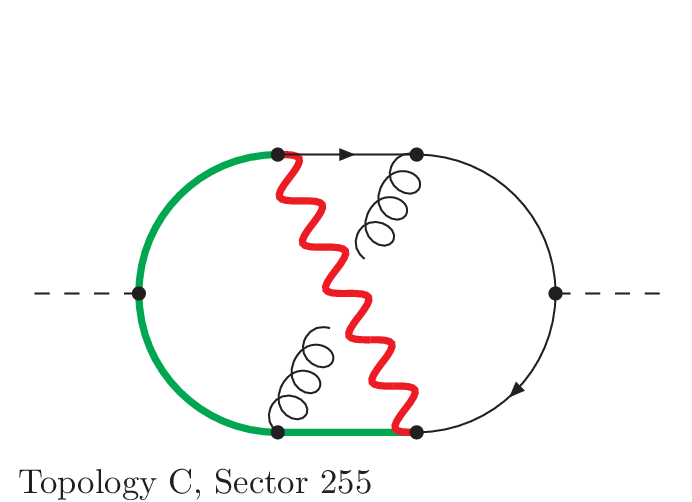}
\hspace*{10mm}
\includegraphics[scale=1.0]{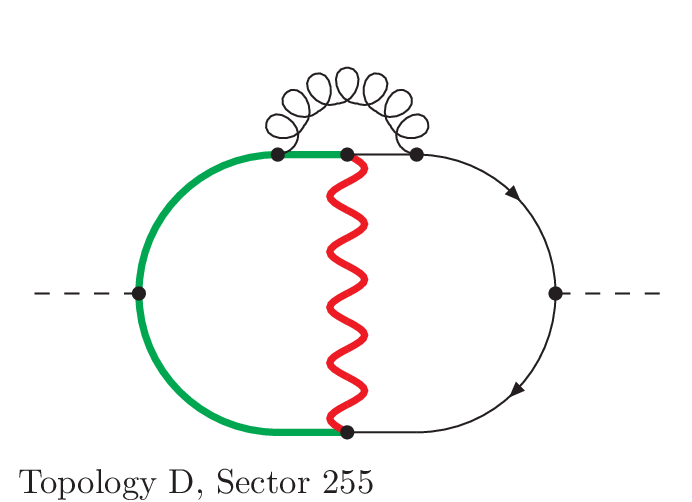}
\end{center}
\caption{
Examples of three-loop Higgs self-energy diagrams containing a top propagator and a $W$-propagator
and being proportional to the product of Yukawa couplings $y_b y_t$.
The internal masses of the propagators are encoded by the colour of the propagators: massless (black), $m_t$ (green), $m_W$ (red).
}
\label{fig_diagrams}
\end{figure}

We consider three-loop Higgs self-energy diagrams of order ${\mathcal O}(\alpha^2 \alpha_s)$
proportional 
to the product of Yukawa couplings $y_b y_t$.
Examples of diagrams 
are shown in fig.~\ref{fig_diagrams}.
The diagrams shown in fig.~\ref{fig_diagrams} have all eight propagators and define the top sectors.
Simpler diagrams, obtained from the ones shown in fig.~\ref{fig_diagrams} by pinching a subset of the propagators,
are not shown.

\begin{figure}
\begin{center}
 \includegraphics[scale=1.0]{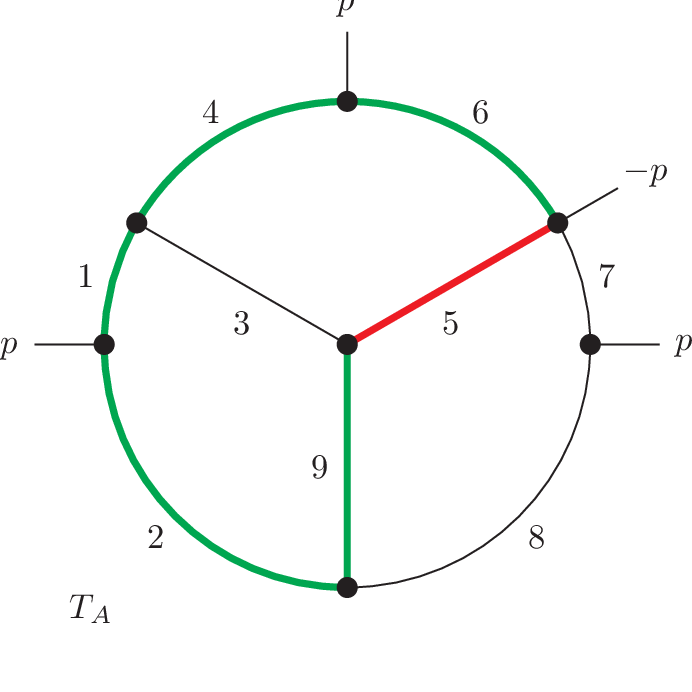}
 \hspace*{10mm}
 \includegraphics[scale=1.0]{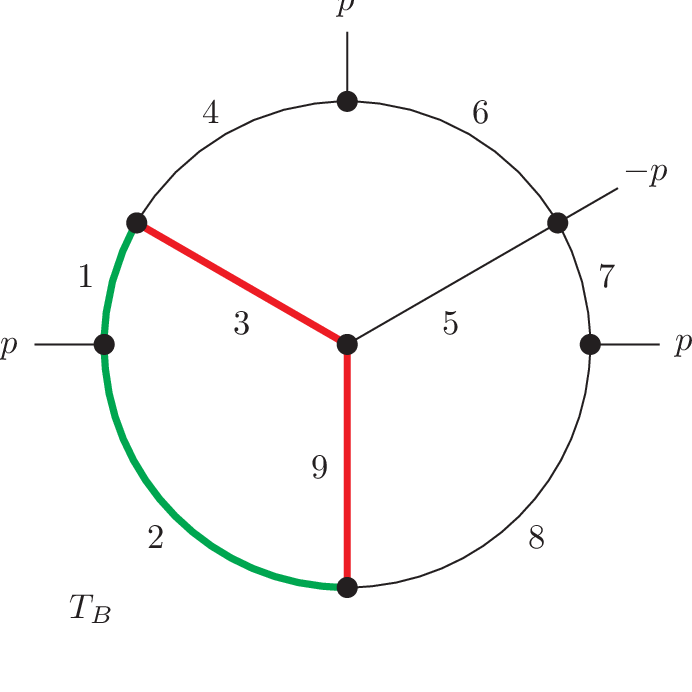}
 \\
 \includegraphics[scale=1.0]{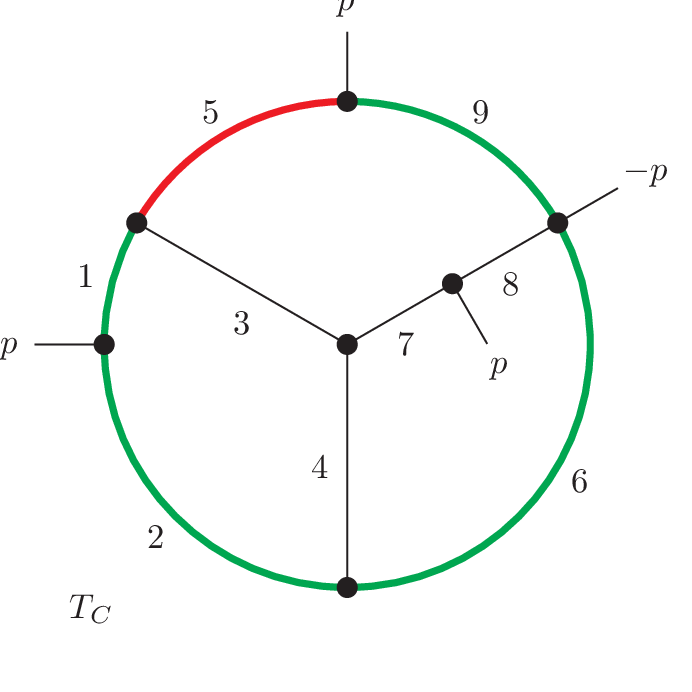}
 \hspace*{10mm}
 \includegraphics[scale=1.0]{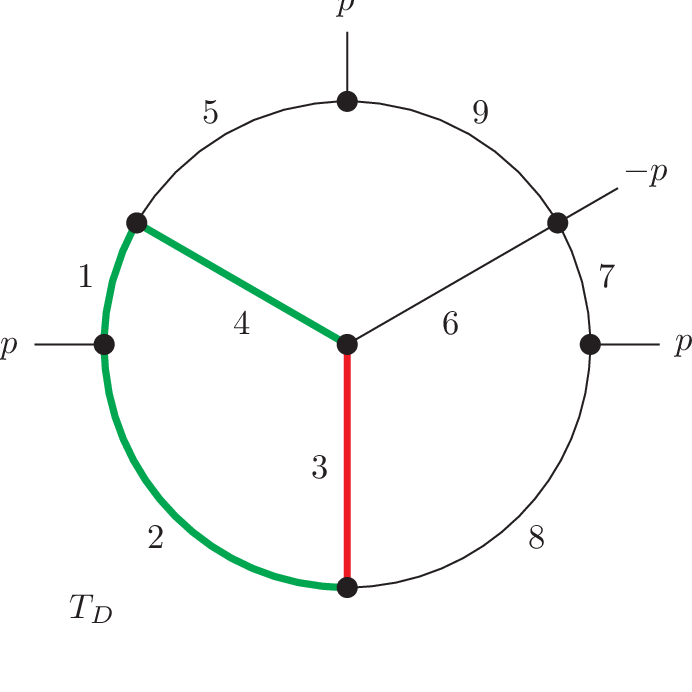}
\end{center}
\caption{
Graphs for the four topologies. The internal masses of the propagators are encoded by the colour of the propagators: massless (black), $m_t$ (green), $m_W$ (red).
}
\label{fig_auxiliary_topologies}
\end{figure}

In this paper we calculate all relevant master integrals.
As we have three independent loop momenta and one independent external momentum, we need nine propagators
in order to be able to express any scalar product involving at least one loop momentum in terms of inverse
propagators.
We therefore consider auxiliary graphs with nine propagators, shown in fig.~\ref{fig_auxiliary_topologies}.
We label these auxiliary graphs topology $A$, $B$, $C$ and $D$
in correspondence with the diagrams shown in fig.~\ref{fig_diagrams}.
 
As a consequence we consider the integrals
\bq
\label{def_integral}
 I_{\nu_1 \nu_2 \nu_3 \nu_4 \nu_5 \nu_6 \nu_7 \nu_8 \nu_9}^X
 & = &
 e^{3 \gamma_E \eps}
 \left(\mu^2\right)^{\nu-\frac{3}{2}D}
 \int \frac{d^Dk_1}{i \pi^{\frac{D}{2}}} \frac{d^Dk_2}{i \pi^{\frac{D}{2}}} \frac{d^Dk_3}{i \pi^{\frac{D}{2}}}
 \prod\limits_{j=1}^9 \frac{1}{ \left(P_j^X\right)^{\nu_j} },
 \nonumber \\
 & & X \in \{A,B,C,D\},
\eq
where $D=4-2\eps$ denotes the number of space-time dimensions, $\gamma_E$ denotes the Euler-Mascheroni constant, 
$\mu$ is an arbitrary scale introduced to render the Feynman integral dimensionless, and the quantity $\nu$ is defined by
\bq
 \nu & = & \sum\limits_{j=1}^9 \nu_j.
\eq
The inverse propagators $P_j^X$ are defined
as follows:
\begin{description}
\item{Topology $A$:}
\begin{align}
 P_1^A
 & =
 -k_1^2 + m_t^2,
 &
 P_2^A
 & =
 -\left(k_1-p\right)^2 + m_t^2,
 &
 P_3^A
 & =
 -\left(k_1+k_2\right)^2,
 \nonumber \\
 P_4^A
 & =
 -k_2^2 + m_t^2,
 &
 P_5^A
 & =
 -\left(k_2+k_3\right)^2 + m_W^2,
 &
 P_6^A
 & =
 -\left(k_2+p\right)^2 + m_t^2,
 \nonumber \\
 P_7^A
 & =
 -k_3^2,
 & 
 P_8^A
 & =
 -\left(k_3-p\right)^2,
 &
 P_9^A
 & =
 -\left(k_1-k_3\right)^2 + m_t^2.
\end{align}
\item{Topology $B$:}
\begin{align}
 P_1^B
 & =
 -k_1^2 + m_t^2,
 &
 P_2^B
 & =
 -\left(k_1-p\right)^2 + m_t^2,
 &
 P_3^B
 & =
 -\left(k_1+k_2\right)^2 + m_W^2,
 \nonumber \\
 P_4^B
 & =
 -k_2^2,
 &
 P_5^B
 & =
 -\left(k_2+k_3\right)^2,
 &
 P_6^B
 & =
 -\left(k_2+p\right)^2,
 \nonumber \\
 P_7^A
 & =
 -k_3^2,
 & 
 P_8^A
 & =
 -\left(k_3-p\right)^2,
 &
 P_9^B
 & =
 -\left(k_1-k_3\right)^2 + m_W^2.
\end{align}
\item{Topology $C$:}
\begin{align}
 P_1^C
 & =
 -k_1^2 + m_t^2,
 &
 P_2^C
 & =
 -\left(k_1-p\right)^2 + m_t^2,
 &
 P_3^C
 & =
 -\left(k_1+k_2\right)^2,
 \nonumber \\
 P_4^C
 & =
 -\left(k_1+k_2-k_3\right)^2,
 &
 P_5^C
 & =
 -k_2^2 + m_W^2,
 &
 P_6^C
 & =
 -\left(k_2-k_3+p\right)^2 + m_t^2,
 \nonumber \\
 P_7^A
 & =
 -k_3^2,
 & 
 P_8^A
 & =
 -\left(k_3-p\right)^2,
 &
 P_9^C
 & =
 -\left(k_2+p\right)^2 + m_t^2.
\end{align}
\item{Topology $D$:}
\begin{align}
 P_1^D
 & =
 -k_1^2 + m_t^2,
 &
 P_2^D
 & =
 -\left(k_1-p\right)^2 + m_t^2,
 &
 P_3^D
 & =
 -\left(k_1-k_3\right)^2 + m_W^2,
 \nonumber \\
 P_4^D
 & =
 -\left(k_1+k_2\right)^2 + m_t^2,
 &
 P_5^D
 & =
 -k_2^2,
 &
 P_6^D
 & =
 -\left(k_2+k_3\right)^2,
 \nonumber \\
 P_7^A
 & =
 -k_3^2,
 & 
 P_8^A
 & =
 -\left(k_3-p\right)^2,
 &
 P_9^D
 & =
 -\left(k_2+p\right)^2.
\end{align}
\end{description}
We introduce the sector id for the Feynman integral $I_{\nu_1 \nu_2 \nu_3 \nu_4 \nu_5 \nu_6 \nu_7 \nu_8 \nu_9}^X$
by
\bq
\label{def_sector_id}
 N_{\mathrm{id}}^X
 \; = \;
 \sum\limits_{j=1}^9 2^{j-1} \Theta\left(\nu_j>0\right).
\eq
With the help of the sector id we may specify all the reductions which we will need:
We have to reduce in every topology
$A-D$ the top sector $255$.
These top sectors correspond to the diagrams shown in fig.~\ref{fig_diagrams}.
The integral reduction gives us in total $105$ master integrals. 
\begin{table}
\begin{center}
\bq
\begin{array}{|l|l|}
 \hline
 \mbox{Topology and sector id} & N_{\mathrm{master}} \\ 
 \hline
 A_{255} & 50 \\ 
 A_{255}+B_{255} & 67 \\ 
 A_{255}+B_{255}+C_{255} & 104 \\ 
 A_{255}+B_{255}+C_{255}+D_{255} & 105 \\ 
 \hline
\end{array}
\eq
\end{center}
\caption{The number of master integrals.
The table gives the cumulative sum, as for example topology $B$ contains master integrals
common to topology $A$ and $B$.
}
\label{table_number_masters}
\end{table}
A detailed list is given in table~\ref{table_number_masters}.
Note that this list gives the cumulative number of master integrals.
This takes into account that certain master integral occur
in more than one topology or sector.
These master integrals are only counted once.
If on the other hand we treat each top sector individually, we find the number of master integrals per top sector
\begin{table}
\begin{center}
\bq
\begin{array}{|l|l|}
 \hline
 \mbox{Topology and sector id} & N_{\mathrm{master}} \\ 
 \hline
 A_{255} & 50 \\
 B_{255} & 32 \\
 C_{255} & 95 \\
 D_{255} & 67 \\
 \hline
\end{array}
\eq
\end{center}
\caption{The number of master integrals for a given top sector, if we treat each top sector individually.
}
\label{table_number_masters_individual}
\end{table}
as shown in table~\ref{table_number_masters_individual}.
In this case the total number of master integrals is higher, as we recompute master integrals, which are identical or are related by
symmetries.
The integral reduction can be carried out with public available computer programs like
{\tt FIRE/Litered} \cite{Smirnov:2008iw,Smirnov:2019qkx,Lee:2012cn,Lee:2013mka},
{\tt Reduze} \cite{Studerus:2009ye,vonManteuffel:2012np},
{\tt Kira} \cite{Maierhoefer:2017hyi,Klappert:2020nbg} or
{\tt FiniteFlow} \cite{Peraro:2016wsq,Peraro:2019svx}.

Let us point out, that it is non-trivial that the total number of master integrals is $105$ and not $106$.
There is one relation not easily found by standard integration-by-parts 
reduction programs\footnote{We thank Johann Usovitsch for contributing this relation.}:
\bq
 I^B_{011011100}
 & = &
 \frac{1}{\left(m_t^2-m_W^2\right)} \left( I^B_{001011100} - I^B_{100011100} \right)
 \nonumber \\
 & &
 - \frac{1}{s} \left( I^A_{01111\left(-1\right)100} - I^A_{011010100} -I^A_{\left(-1\right)11110100} \right)
 \nonumber \\
 & &
 - \frac{3}{s\left(m_t^2-m_W^2\right)} \left( I^A_{011\left(-1\right)10100} - I^A_{01101\left(-1\right)100} \right).
\eq
Without this relation we may end up with a non-integrable system.

The Feynman parameter representations for the four topologies
are given by
\bq
 I_{\nu_1 \nu_2 \nu_3 \nu_4 \nu_5 \nu_6 \nu_7 \nu_8 \nu_9}^X
 & = &
 e^{3 \gamma_E \eps}
 \frac{\Gamma(\nu-\frac{3}{2}D)}{\prod\limits_{j=1}^{9}\Gamma(\nu_j)}
 \;
 \int\limits_{x_i \ge 0}
 d^9x
 \;
 \delta\left(1-\sum\limits_{j=1}^9 x_j\right)
 \; 
 \left( \prod\limits_{j=1}^{9} x_j^{\nu_j-1} \right)\,\frac{{\mathcal U}_X^{\nu-2D}}
 {{\mathcal F}_X^{\nu-\frac{3}{2}D}},
\eq
where ${\mathcal U}_X$ and ${\mathcal F}_X$ denote the first and second graph polynomial, respectively.
The first graph polynomials for the individual topologies are given by
\bq
 {\mathcal U}_A
 & = & {\mathcal U}_B
 \; = \;
  x_{12}\,x_3\,x_5+x_{12}\,x_3\,x_{78}+x_{12}\,x_3\,x_9+x_{12}\,x_{46}\,x_5+x_{12}\,x_{46}\,x_{78}
 +x_{12}\,x_{46}\,x_9+x_{12}\,x_5\,x_{78}
 \nonumber \\
 & &
 +x_{12}\,x_5\,x_9+x_3\,x_{46}\,x_5+x_3\,x_{46}\,x_{78}+x_3\,x_{46}\,x_9+x_3\,x_5\,x_{78}+x_3\,x_{78}\,x_9
 +x_{46}\,x_5\,x_9+x_{46}\,x_{78}\,x_9
 \nonumber \\
 & &
 +x_5\,x_{78}\,x_9,
 \nonumber \\
 {\mathcal U}_C
 & = & 
 x_{12}\,x_3\,x_4+x_{12}\,x_3\,x_6+x_{12}\,x_3\,x_{78}+x_{12}\,x_4\,x_{59}+x_{12}\,x_4\,x_{78}+x_{12}\,x_{59}\,x_6
 +x_{12}\,x_{59}\,x_{78}
 \nonumber \\
 & &
 +x_{12}\,x_6\,x_{78}+x_3\,x_4\,x_{59}+x_3\,x_4\,x_6+x_3\,x_{59}\,x_6+x_3\,x_{59}\,x_{78}+x_3\,x_6\,x_{78}
 +x_4\,x_{59}\,x_6+x_4\,x_{59}\,x_{78}
 \nonumber \\
 & &
 +x_4\,x_6\,x_{78},
 \nonumber \\
 {\mathcal U}_D
 & = & 
 x_{12}\,x_3\,x_4+x_{12}\,x_3\,x_{59}+x_{12}\,x_3\,x_6+x_{12}\,x_4\,x_6+x_{12}\,x_4\,x_{78}+x_{12}\,x_{59}\,x_6
 +x_{12}\,x_{59}\,x_{78}
 \nonumber \\
 & &
 +x_{12}\,x_6\,x_{78}+x_3\,x_4\,x_{59}+x_3\,x_4\,x_{78}+x_3\,x_{59}\,x_6+x_3\,x_{59}\,x_{78}
 +x_3\,x_6\,x_{78}+x_4\,x_{59}\,x_6+x_4\,x_{59}\,x_{78}
 \nonumber \\
 & &
 +x_4\,x_6\,x_{78},
\eq
where we used the abbreviation $x_{ij} = x_i +x_j$.

\subsection{Dimensional shift relations and differential equations}

Let us introduce an operator ${\bf i}^+$, which raises the power of the propagator $i$ by one and multiplies by $\nu_i$, e.g.
\bq
 {\bf 1}^+ I_{\nu_1 \nu_2 \nu_3 \nu_4 \nu_5 \nu_6 \nu_7 \nu_8 \nu_9}^X & = &
 \nu_1 \cdot I_{(\nu_1+1) \nu_2 \nu_3 \nu_4 \nu_5 \nu_6 \nu_7 \nu_8 \nu_9}^X.
\eq
In addition we define two operators ${\bf D}^\pm$,
which shift the dimension of space-time by two through
\bq
 {\bf D}^\pm I_{\nu_1 \nu_2 \nu_3 \nu_4 \nu_5 \nu_6 \nu_7 \nu_8 \nu_9}^X\left( D \right) & = &
 I_{\nu_1 \nu_2 \nu_3 \nu_4 \nu_5 \nu_6 \nu_7 \nu_8 \nu_9}^X\left( D\pm 2 \right).
\eq
The dimensional shift relations 
read \cite{Tarasov:1996br,Tarasov:1997kx}
\bq
\label{dim_shift_eq}
\lefteqn{
 {\bf D}^- I_{\nu_1 \nu_2 \nu_3 \nu_4 \nu_5 \nu_6 \nu_7 \nu_8 \nu_9}^X\left(D\right)
 = } &  &
 \\
 & &
 {\mathcal U}_X\left( {\bf 1}^+, {\bf 2}^+, {\bf 3}^+, {\bf 4}^+, {\bf 5}^+ , {\bf 6}^+, {\bf 7}^+ , {\bf 8}^+ , {\bf 9}^+ \right)
 I_{\nu_1 \nu_2 \nu_3 \nu_4 \nu_5 \nu_6 \nu_7 \nu_8 \nu_9}^X\left(D\right).
 \nonumber
\eq
We denote by $J^X_{m_i^2}$ the set of internal edge indices, which correspond to the propagation of a particle with mass $m_i$.
For example
\bq
 J^A_{m_t^2} & = & \left\{ 1, 2, 4, 6, 9 \right\}.
\eq
With this notation the differential equations read
\bq
 \mu^2 \frac{\partial}{\partial m_t^2} I^X_{\nu_1 \nu_2 \nu_3 \nu_4 \nu_5 \nu_6 \nu_7 \nu_8 \nu_9}
 & = &
 - \sum\limits_{j \in J^X_{m_t^2}} {\bf j}^+ I^X_{\nu_1 \nu_2 \nu_3 \nu_4 \nu_5 \nu_6 \nu_7 \nu_8 \nu_9},
 \nonumber \\
 \mu^2 \frac{\partial}{\partial m_W^2} I^X_{\nu_1 \nu_2 \nu_3 \nu_4 \nu_5 \nu_6 \nu_7 \nu_8 \nu_9}
 & = &
 - \sum\limits_{j \in J^X_{m_W^2}} {\bf j}^+ I^X_{\nu_1 \nu_2 \nu_3 \nu_4 \nu_5 \nu_6 \nu_7 \nu_8 \nu_9},
 \nonumber \\
 \mu^2 \frac{\partial}{\partial s} I^X_{\nu_1 \nu_2 \nu_3 \nu_4 \nu_5 \nu_6 \nu_7 \nu_8 \nu_9}
 & = &
   \left( \frac{3}{2} D -\nu \right) \frac{\mu^2}{s} I^X_{\nu_1 \nu_2 \nu_3 \nu_4 \nu_5 \nu_6 \nu_7 \nu_8 \nu_9}
   + \frac{m_t^2}{s} \sum\limits_{j \in J^X_{m_t^2}} {\bf j}^+ I^X_{\nu_1 \nu_2 \nu_3 \nu_4 \nu_5 \nu_6 \nu_7 \nu_8 \nu_9}
 \nonumber \\
 & &
   + \frac{m_W^2}{s} \sum\limits_{j \in J^X_{m_W^2}} {\bf j}^+ I^X_{\nu_1 \nu_2 \nu_3 \nu_4 \nu_5 \nu_6 \nu_7 \nu_8 \nu_9}.
\eq
For the derivative with respect to $s$ we used the scaling relation. This has the advantage that on the right-hand side at most
one raising operator occurs.
This minimises the integral reductions we have to compute.

\subsection{Variables}

In the following we will set
\bq
 \mu^2 & = & m_t^2.
\eq
The master integrals depend then kinematically on two dimensionless quantities.
We set
\bq
 v \; = \; \frac{p^2}{m_t^2},
 & &
 w \; = \; \frac{m_W^2}{m_t^2}.
\eq
We will encounter the following two square roots
\bq
 r_1
 \; = \;
 \sqrt{-v\left(4-v\right)},
 & &
 r_2
 \; = \;
 \sqrt{\lambda\left(v,w,1\right)}.
\eq
The K\"allen function is defined by
\bq
 \lambda(x,y,z)
 & = &
 x^2 + y^2 + z^2 - 2 x y - 2 y z - 2 z x.
\eq
The variable transformation from $(v,w)$ to $(x,y)$, where the latter variables
are defined by \cite{Chaubey:2019lum,Besier:2018jen,Besier:2019kco}
\bq
\label{variable_trafo}
 v \; = \; - \frac{\left(1-x\right)^2}{x},
 & &
 w \; = \; \frac{\left(1-y+2xy\right)\left(x-2y+xy\right)}{x\left(1-y^2\right)},
\eq
rationalises the square roots $r_1$ and $r_2$:
\bq
 r_1 \; = \; \frac{1-x^2}{x},
 & &
 r_2 \; = \; \frac{\left(1-x\right)\left[\left(1-y\right)^2+x\left(1+y\right)^2\right]}{x\left(1-y^2\right)}.
\eq
The inverse transformation is given by
\bq
\label{variable_trafo_inv}
 x \; = \; \frac{1}{2} \left( 2-v - r_1 \right),
 & & 
 y \; = \;
 \frac{r_2 - r_1}{1-w+2v},
\eq
such that $x=0$ corresponds to $v=\infty$ and $y=0$ corresponds to $w=1$.
We note that the point $(v,w)=(0,1)$ is blown up in $(x,y)$-space to the hypersurface $x=1$.
This is convenient, as it allows us to obtain the results for the master integrals by integrating the differential equation in
$x$ with constant $y$. We make a final change of coordinates and we introduce
\bq
 x' & = & 1 - x.
\eq
This transforms the starting point of the integration from $x=1$ to $x'=0$ and is helpful for the conversion 
to multiple polylogarithms.


\section{Master integrals}
\label{sect:masters}

In this section we define master integrals of uniform weight.
This is the main result of this paper.

We have $105$ master integrals and $59$ master topologies (or sectors).
These are summarised in table~\ref{table_master_integrals_A} and \ref{table_master_integrals_BCD}.
We denote the master integrals of uniform weight by $J_1$-$J_{105}$.

Existing deterministic algorithms \cite{Moser:1959,Lee:2014ioa,Lee:2017oca,Prausa:2017ltv,Gituliar:2017vzm,Lee:2020zfb}
for the construction of master integrals of uniform weight cannot be used for problems of this complexity.
Instead, we revert to heuristic methods.
We first note that given an educated guess for a basis of master integrals of uniform weight, it is easy to verify/falsify 
if the proposed basis has the desired properties: One simply computes the differential equation in this basis.
For the educated guess we study the maximal cuts in the loop-by-loop Baikov representation \cite{Baikov:1996iu,Frellesvig:2017aai}
and construct integrands, which give to leading order in $\eps$ constants when integrated over a basis of independent contours \cite{Cachazo:2008vp,Arkani-Hamed:2010pyv}.
In a second step, possible contributions from sub-topologies are fixed by an ansatz.
A pedagogical description of this method can be found in chapter $16$ of \cite{Weinzierl:2022eaz}.

For all these master integrals we find compact expressions in terms of pre-canonical Feynman integrals.
It is worth noting that all prefactors are polynomials in $\eps,v,w,r_1,r_2$, i.e. no denominators occur.
The master integrals $J_1$-$J_{105}$ are defined by
\bq
\label{def_master_integrals}
 J_{1}
 & = &
 \eps^3 \; {\bf D}^- I^A_{100110000},
 \nonumber \\
 J_{2}
 & = &
 \eps^3 \; r_1 \; {\bf D}^- I^A_{110110000},
 \nonumber \\
 J_{3}
 & = &
 \eps^3 \; r_1 \; {\bf D}^- I^A_{011110000},
 \nonumber \\
 J_{4}
 & = &
 \eps^3 \; {\bf D}^- I^A_{\left(-1\right)11110000},
 \nonumber \\
 J_{5}
 & = &
 \eps^3 \; \left(1-w\right) \; {\bf D}^- I^A_{101010100},
 \nonumber \\
 J_{6}
 & = &
 \eps^3 \; {\bf D}^- I^A_{101\left(-1\right)10100},
 \nonumber \\
 J_{7}
 & = &
 \eps^3 \; r_2 \; {\bf D}^- I^A_{011010100},
 \nonumber \\
 J_{8}
 & = &
 \eps^3 \; \left[ {\bf D}^- I^A_{011\left(-1\right)10100} - \left(1-w\right) \; {\bf D}^- I^A_{011010100} \right],
 \nonumber \\
 J_{9}
 & = &
 \eps^3 \; {\bf D}^- I^A_{01101\left(-1\right)100},
 \nonumber \\
 J_{10}
 & = &
 \eps^3 \; \left[ {\bf D}^- I^A_{0110101\left(-1\right)0} + v \; {\bf D}^- I^A_{011010100} \right],
 \nonumber \\
 J_{11}
 & = &
 \eps^3 \; \left(1-w\right) \; {\bf D}^- I^A_{100110100},
 \nonumber \\
 J_{12}
 & = &
 \eps^3 \; r_2 \; {\bf D}^- I^A_{100011100},
 \nonumber \\
 J_{13}
 & = &
 \eps^3 \; \left[ {\bf D}^- I^A_{100\left(-1\right)11100} - \left(1-w\right) \; {\bf D}^- I^A_{100011100} \right],
 \nonumber \\
 J_{14}
 & = &
 \eps^3 \; \left[ {\bf D}^- I^A_{1000111\left(-1\right)0} + v \; {\bf D}^- I^A_{100011100} \right],
 \nonumber \\
 J_{15}
 & = &
 \eps^3 \; v \; {\bf D}^- I^A_{100100110},
 \nonumber \\
 J_{16}
 & = &
 \eps^3 \; v \; {\bf D}^- I^A_{100010110},
 \nonumber \\
 J_{17}
 & = &
 \eps^3 \; v\left(4-v\right) \; {\bf D}^- I^A_{110111000},
 \nonumber \\
 J_{18}
 & = &
 \eps^3 \; r_1 \; \left[ {\bf D}^- I^A_{111\left(-1\right)10100} - \left(1-w\right) \; {\bf D}^- I^A_{111010100}\right],
 \nonumber \\
 J_{19}
 & = &
 \eps^3 \; r_1 \; \left[ {\bf D}^- I^A_{111\left(-1\right)10100} - {\bf D}^- I^A_{011010100}\right],
 \nonumber \\
 J_{20}
 & = &
 \eps^3 \; \left(1-w\right) r_1 \; {\bf D}^- I^A_{110110100},
 \nonumber \\
 J_{21}
 & = &
 \eps^3 \; r_1 \left[ \left(1-w\right) \; {\bf D}^- I^A_{011110100} - {\bf D}^- I^A_{011010100} \right],
 \nonumber \\
 J_{22}
 & = &
 \eps^3 \; \left(1-w\right) \left[ {\bf D}^- I^A_{\left(-1\right)11110100} - {\bf D}^- I^A_{011010100} \right],
 \nonumber \\
 J_{23}
 & = &
 \eps^3 \; \left(1-w\right) \; {\bf D}^- I^A_{01111\left(-1\right)100},
 \nonumber \\
 J_{24}
 & = &
 \eps^3 \; r_1 r_2 \; {\bf D}^- I^A_{110011100},
 \nonumber \\
 J_{25}
 & = &
 \eps^3 \; r_1 \left[ {\bf D}^- I^A_{110\left(-1\right)11100} - \left(1-w\right) \; {\bf D}^- I^A_{110011100} \right],
 \nonumber \\
 J_{26}
 & = &
 \eps^3 \; r_1 \left[ {\bf D}^- I^A_{1100111\left(-1\right)0} + v \; {\bf D}^- I^A_{110011100} \right],
 \nonumber \\
 J_{27}
 & = &
 \eps^3 \; r_2 \; {\bf D}^- I^A_{101\left(-1\right)11100},
 \nonumber \\
 J_{28}
 & = &
 \eps^3 \; r_1 \left[ {\bf D}^- I^A_{101\left(-1\right)11100} - \left(1-w\right) \; {\bf D}^- I^A_{101011100} \right],
 \nonumber \\
 J_{29}
 & = &
 \eps^3 \; \left[ {\bf D}^- I^A_{101\left(-2\right)11100} - \left(1-w\right) \; {\bf D}^- I^A_{101\left(-1\right)11100} \right],
 \nonumber \\
 J_{30}
 & = &
 \eps^3 \; \left(1-v\right) \left[ {\bf D}^- I^A_{101\left(-1\right)11100} + {\bf D}^- I^A_{10101110\left(-1\right)} - \left(1-w\right) \; {\bf D}^- I^A_{101011100} \right],
 \nonumber \\
 J_{31}
 & = &
 \eps^3 \; r_1 \left[ \left(1-w\right) \; {\bf D}^- I^A_{100111100} - {\bf D}^- I^A_{100011100} \right],
 \nonumber \\
 J_{32}
 & = &
 \eps^3 \; v r_1 \;  {\bf D}^- I^A_{110100110},
 \nonumber \\
 J_{33}
 & = &
 \eps^3 \; v r_1 \; {\bf D}^- I^A_{011100110},
 \nonumber \\
 J_{34}
 & = &
 \eps^3 \; v \; {\bf D}^- I^A_{\left(-1\right)11100110},
 \nonumber \\
 J_{35}
 & = &
 \eps^3 \; v r_1 \; {\bf D}^- I^A_{110010110},
 \nonumber \\
 J_{36}
 & = &
 \eps^3 \; \left(1-w\right) \left[ v \; {\bf D}^- I^A_{101010110} + {\bf D}^- I^A_{011010100} \right],
 \nonumber \\
 J_{37}
 & = &
 \eps^4 \left(1-2\eps\right) \; v \; I^A_{111021100},
 \nonumber \\
 J_{38}
 & = &
 \eps^4 \left(1-2\eps\right) \; v \; I^A_{111011200},
 \nonumber \\
 J_{39}
 & = &
 \eps^3 \; v\left(4-v\right) \left[ \left(1-w\right) \; {\bf D}^- I^A_{110111100} - {\bf D}^- I^A_{110011100} \right],
 \nonumber \\
 J_{40}
 & = &
 \eps^3 v r_1 \left[ {\bf D}^- I^A_{11101011(-1)} - \left(1-w\right) \; {\bf D}^- I^A_{111010110} \right],
 \nonumber \\
 J_{41}
 & = &
 \eps^4 \left(1-2\eps\right) \; v \; I^A_{112010110},
 \nonumber \\
 J_{42}
 & = &
 \eps^4 \left(1-2\eps\right) \; v \; I^A_{111020110},
 \nonumber \\
 J_{43}
 & = &
 \eps^4 \left(1-2\eps\right) \; v \; I^A_{021110110},
 \nonumber \\
 J_{44}
 & = &
 \eps^4 \left(1-2\eps\right) \; v \; I^A_{012110110},
 \nonumber \\
 J_{45}
 & = &
 \eps^3 \; v^2\left(4-v\right) \; {\bf D}^- I^A_{110101110},
 \nonumber \\
 J_{46}
 & = &
 \eps^4 \left(1-\eps\right) \left(1-2\eps\right) \; v \; I^A_{100111110},
 \nonumber \\
 J_{47}
 & = &
 \eps^5 \left(1-2\eps\right) \; v \; I^A_{111110110},
 \nonumber \\
 J_{48}
 & = &
 \eps^4 \left(1-2\eps\right) \; v \left(1-w\right) \; I^A_{111120110},
 \nonumber \\
 J_{49}
 & = &
 \eps^4 \left(1-2\eps\right) \; v \left[ I^A_{211110110} + \frac{1}{2}\left(2-v+r_1\right) \; I^A_{121110110} \right],
 \nonumber \\
 J_{50}
 & = &
 \eps^4 \left(1-2\eps\right) \; v r_1 \; I^A_{210111110},
 \nonumber \\
 J_{51}
 & = &
 \eps^3 \; v \; {\bf D}^- I^B_{100011100},
 \nonumber \\
 J_{52}
 & = &
 \eps^3 \; v \; {\bf D}^- I^B_{001011100},
 \nonumber \\
 J_{53}
 & = &
 \eps^3 \; v r_1 \; {\bf D}^- I^B_{110011100},
 \nonumber \\
 J_{54}
 & = &
 \eps^3 \; r_2 \; {\bf D}^- I^B_{101\left(-1\right)11100},
 \nonumber \\
 J_{55}
 & = &
 \eps^3 \; \left(1-w\right) \left[ {\bf D}^- I^B_{101\left(-1\right)11100} + v \; {\bf D}^- I^B_{101011100} \right],
 \nonumber \\
 J_{56}
 & = &
 \eps^3 \left[ {\bf D}^- I^B_{101\left(-2\right)11100} + v \; {\bf D}^- I^B_{101\left(-1\right)11100} \right],
 \nonumber \\
 J_{57}
 & = &
 \eps^3 \; v r_2 \; {\bf D}^- I^B_{011100110},
 \nonumber \\
 J_{58}
 & = &
 \eps^3 \; v \left[ {\bf D}^- I^B_{\left(-1\right)11100110} - \left(1-w\right) \; {\bf D}^- I^B_{011100110} \right],
 \nonumber \\
 J_{59}
 & = &
 \eps^3 \; v \; {\bf D}^- I^B_{0111\left(-1\right)0110},
 \nonumber \\
 J_{60}
 & = &
 \eps^3 \; v^2 \; {\bf D}^- I^B_{100101110},
 \nonumber \\
 J_{61}
 & = &
 \eps^3 \; v^2 \; {\bf D}^- I^B_{001101110},
 \nonumber \\
 J_{62}
 & = &
 \eps^4 \left(1-2\eps\right) \; v \; I^B_{111021100},
 \nonumber \\
 J_{63}
 & = &
 \eps^3 \; r_1
   \left\{ \left(1-w\right) \; {\bf D}^- I^B_{111\left(-1\right)11100}
          - {\bf D}^- I^B_{101\left(-1\right)11100}
 \right\},
 \nonumber \\
 J_{64}
 & = &
 \eps^3 \; v r_1
   \left\{ \left(1-w\right) \; {\bf D}^- I^B_{111100110}
          - {\bf D}^- I^B_{011100110}
 \right\},
 \nonumber \\
 J_{65}
 & = &
 \eps^3 \; v^2 r_1 \; {\bf D}^- I^B_{110101110},
 \nonumber \\
 J_{66}
 & = &
 \eps^5 \left(1-2\eps\right) \; v \; I^B_{111110110},
 \nonumber \\
 J_{67}
 & = &
 \eps^4 \left(1-2\eps\right)^2 \; v \; I^B_{111101110},
 \nonumber \\
 J_{68}
 & = &
 \eps^3 \; r_1 \; {\bf D}^- I^C_{101001100},
 \nonumber \\
 J_{69}
 & = &
 \eps^3 \; {\bf D}^- I^C_{1\left(-1\right)1001100},
 \nonumber \\
 J_{70}
 & = &
 \eps^3 \; {\bf D}^- I^C_{10100110\left(-1\right)},
 \nonumber \\
 J_{71}
 & = &
 \eps^2 \left(1+4\eps\right) \; {\bf D}^- I^C_{011001100},
 \nonumber \\
 J_{72}
 & = &
 \eps^3 r_1 \; {\bf D}^- I^C_{11100110\left(-1\right)},
 \nonumber \\
 J_{73}
 & = &
 \eps^3 \; r_2 \left[ \left(1-w\right) \; {\bf D}^- I^C_{101011100}
   + {\bf D}^- I^C_{101001100}
 \right],
 \nonumber \\
 J_{74}
 & = &
 \eps^3 \; \left(1-w\right) \; {\bf D}^- I^C_{10101110\left(-1\right)},
 \nonumber \\
 J_{75}
 & = &
 \eps^3 \; \left(1-w\right) \left[ {\bf D}^- I^C_{1010111\left(-1\right)0} + v \; {\bf D}^- I^C_{101011100} \right],
 \nonumber \\
 J_{76}
 & = &
 \eps^3 \; r_2 \; {\bf D}^- I^C_{01101110\left(-1\right)},
 \nonumber \\
 J_{77}
 & = &
 \eps^3 \; {\bf D}^- I^C_{01101110\left(-2\right)},
 \nonumber \\
 J_{78}
 & = &
 \eps^3 \; \left(w-v\right) \left\{ {\bf D}^- I^C_{011\left(-1\right)11100}
  - 2 \; {\bf D}^- I^A_{100011100}
 \right\},
 \nonumber \\
 J_{79}
 & = &
 \eps^3 \; r_1  
   \left\{
          2 \; {\bf D}^- I^C_{1\left(-1\right)1101010}
          + v \; {\bf D}^- I^C_{101101010} 
          - {\bf D}^- I^C_{011001100}
          - 2 \; {\bf D}^- I^C_{101100010}
 \right\},
 \nonumber \\
 J_{80}
 & = &
 \eps^3 \; v \; \left[
  2 \left(1-3\eps\right) \left( I^C_{202101010} - \eps \; I^C_{102101010} \right)
  + \eps \left(1-2\eps\right) I^C_{101201010}
 \right],
 \nonumber \\
 J_{81}
 & = &
 \eps^3 \left(1-w\right) 
  \left[ \left(1-w\right) \; {\bf D}^- I^C_{101011010} + {\bf D}^- I^C_{011001100} \right],
 \nonumber \\
 J_{82}
 & = &
 \eps^3 \; r_1
 \left\{ \left(1-w\right) \; {\bf D}^- I^C_{11101110\left(-1\right)} - {\bf D}^- I^C_{01101110\left(-1\right)}\right\},
 \nonumber \\
 J_{83}
 & = &
 \eps^4 \left(1-2\eps\right) \; v \; I^C_{111011200},
 \nonumber \\
 J_{84}
 & = &
 \eps^4 \left(1-2\eps\right) \; v \; I^C_{111012100},
 \nonumber \\
 J_{85}
 & = &
 \eps^4 \left(1-2\eps\right) \; v \; I^C_{112101010},
 \nonumber \\
 J_{86}
 & = &
 \eps^3 \; \left(1-w\right) r_1 \left[
  \left(1-w\right) \; {\bf D}^- I^C_{111011010}
  - {\bf D}^- I^C_{101011100}
  + {\bf D}^- I^C_{111001100}
 \right],
 \nonumber \\
 J_{87}
 & = &
 \eps^4 \left(1-2\eps\right)
   \left[
          I^C_{101111010} - \frac{1}{2} \left(1-w\right) \; I^C_{201111010} - \frac{3}{2} \; I^C_{101211(-1)10}
          + \frac{1}{4} v \; I^C_{101201010}
 \right. \nonumber \\
 & & \left.
          - \frac{1}{2} \; I^C_{201101010}
          + \frac{1}{2} \; I^C_{200111010} + \frac{3}{2} \; I^C_{100211010}
   \right] 
 + \frac{\eps^3}{2} \left(1-3\eps\right) \; v \left[ I^C_{202101010} - \eps \; I^C_{102101010} \right],
 \nonumber \\
 J_{88}
 & = &
 \eps^4 \left(1-2\eps\right) \; v \; I^C_{101211010},
 \nonumber \\
 J_{89}
 & = &
 \eps^4 \left(1-2\eps\right) \; v \; I^C_{012111010},
 \nonumber \\
 J_{90}
 & = &
 \eps^4 \left(1-2\eps\right) \; v \; I^C_{112001110},
 \nonumber \\
 J_{91}
 & = &
 \eps^4 \left(1-2\eps\right) \; v \; I^C_{111002110},
 \nonumber \\
 J_{92}
 & = &
 \eps^2 \; r_1 \left[ 
                      \left(1-2\eps\right)^2 \; v \; I^C_{211001210}
                      -\eps  \left(1-2\eps\right) \; I^C_{111002110}
                      - I^C_{202002100}
                      - v \; I^A_{210200210}
               \right],
 \nonumber \\
 J_{93}
 & = &
 \eps^4 \left(1-2\eps\right) \; v \; I^C_{021011110},
 \nonumber \\
 J_{94}
 & = &
 \eps^4 \left(1-2\eps\right) \; v \; I^C_{012011110},
 \nonumber \\
 J_{95}
 & = &
 \eps^5 \left(1-2\eps\right) \; v \; I^C_{111111010},
 \nonumber \\
 J_{96}
 & = &
 \eps^4 \left(1-2\eps\right) \; v w \; I^C_{111121010},
 \nonumber \\
 J_{97}
 & = &
 \eps^4 \left(1-2\eps\right) \; v \left(1-w\right) \; I^C_{111111020},
 \nonumber \\
 J_{98}
 & = &
 \eps^4 \left(1-2\eps\right) \; r_1 \left( I^C_{211111010} + I^C_{121111010} + I^C_{111112010} + w \; I^C_{111121010} - 4 \eps \; I^C_{111111010} \right),
 \nonumber \\
 J_{99}
 & = &
 \eps^5 \left(1-2\eps\right) \; v \; I^C_{111011110},
 \nonumber \\
 J_{100}
 & = &
 \eps^4 \left(1-2\eps\right) \; v \left(1-w\right) \; I^C_{111012110},
 \nonumber \\
 J_{101}
 & = &
 \eps^4 \left(1-2\eps\right) \; v \left(1-w\right) \; I^C_{112011110},
 \nonumber \\
 J_{102}
 & = &
 \eps^4 \left(1-2\eps\right) \; v r_1 \; I^C_{121011110},
 \nonumber \\
 J_{103}
 & = &
 \eps^5 \left(1-2\eps\right)  \; v \; I^C_{101111110},
 \nonumber \\
 J_{104}
 & = &
 \eps^5 \left(1-2\eps\right)  \; v \left(1-v-w\right) \; I^C_{111111110},
 \nonumber \\
 J_{105}
 & = &
 \eps^4 \left(1-2\eps\right)  \; v w \; I^D_{111201110}.
\eq
\begin{table}[!htbp]
\begin{center}
\begin{tabular}{|c|r|r|l|l|}
\hline
 topology & block & sector & propagators & master integrals \\
\hline
\hline
 $A$ & $1$ & $25$ & $1,4,5$ & $J_{1}$ \\
     & $2$ & $27$ & $1,2,4,5$ & $J_{2}$ \\
     & $3$ & $30$ & $2,3,4,5$ & $J_{3},J_{4}$ \\
     & $4$ & $85$ & $1,3,5,7$ & $J_{5},J_{6}$ \\
     & $5$ & $86$ & $2,3,5,7$ & $J_{7},J_{8},J_{9},J_{10}$ \\
     & $6$ & $89$ & $1,4,5,7$ & $J_{11}$ \\
     & $7$ & $113$ & $1,5,6,7$ & $J_{12},J_{13},J_{14}$ \\
     & $8$ & $201$ & $1,4,7,8$ & $J_{15}$ \\
     & $9$ & $209$ & $1,5,7,8$ & $J_{16}$ \\
     & $10$ & $59$ & $1,2,4,5,6$ & $J_{17}$ \\
     & $11$ & $87$ & $1,2,3,5,7$ & $J_{18},J_{19}$ \\
     & $12$ & $91$ & $1,2,4,5,7$ & $J_{20}$ \\
     & $13$ & $94$ & $2,3,4,5,7$ & $J_{21},J_{22},J_{23}$ \\
     & $14$ & $115$ & $1,2,5,6,7$ & $J_{24},J_{25},J_{26}$ \\
     & $15$ & $117$ & $1,3,5,6,7$ & $J_{27},J_{28},J_{29},J_{30}$ \\
     & $16$ & $121$ & $1,4,5,6,7$ & $J_{31}$ \\
     & $17$ & $203$ & $1,2,4,7,8$ & $J_{32}$ \\
     & $18$ & $206$ & $2,3,4,7,8$ & $J_{33},J_{34}$ \\
     & $19$ & $211$ & $1,2,5,7,8$ & $J_{35}$ \\
     & $20$ & $213$ & $1,3,5,7,8$ & $J_{36}$ \\
     & $21$ & $119$ & $1,2,3,5,6,7$ & $J_{37},J_{38}$ \\
     & $22$ & $123$ & $1,2,4,5,6,7$ & $J_{39}$ \\
     & $23$ & $215$ & $1,2,3,5,7,8$ & $J_{40},J_{41},J_{42}$ \\
     & $24$ & $222$ & $2,3,4,5,7,8$ & $J_{43},J_{44}$ \\
     & $25$ & $235$ & $1,2,4,6,7,8$ & $J_{45}$ \\
     & $26$ & $249$ & $1,4,5,6,7,8$ & $J_{46}$ \\
     & $27$ & $223$ & $1,2,3,4,5,7,8$ & $J_{47},J_{48},J_{49}$ \\
     & $28$ & $251$ & $1,2,4,5,6,7,8$ & $J_{50}$ \\
\hline
\end{tabular}
\end{center}
\caption{Overview of the set of master integrals.
The first column denotes the topology, 
the second column labels consecutively the sectors,
the third column gives the sector id (defined in eq.~(\ref{def_sector_id})),
the fourth column lists the propagators with positive exponent,
the fifth column lists the master integrals in the basis $\vec{J}$.
}
\label{table_master_integrals_A}
\end{table}
\begin{table}[!htbp]
\begin{center}
\begin{tabular}{|c|r|r|l|l|}
\hline
 topology & block & sector & propagators & master integrals \\
\hline
\hline
 $B$ & $29$ & $113$ & $1,5,6,7$ & $J_{51}$ \\
     & $30$ & $116$ & $3,5,6,7$ & $J_{52}$ \\
     & $31$ & $115$ & $1,2,5,6,7$ & $J_{53}$ \\
     & $32$ & $117$ & $1,3,5,6,7$ & $J_{54},J_{55},J_{56}$ \\
     & $33$ & $206$ & $2,3,4,7,8$ & $J_{57},J_{58},J_{59}$ \\
     & $34$ & $233$ & $1,4,6,7,8$ & $J_{60}$ \\
     & $35$ & $236$ & $3,4,6,7,8$ & $J_{61}$ \\
     & $36$ & $119$ & $1,2,3,5,6,7$ & $J_{62},J_{63}$ \\
     & $37$ & $207$ & $1,2,3,4,7,8$ & $J_{64}$ \\
     & $38$ & $235$ & $1,2,4,6,7,8$ & $J_{65}$ \\
     & $39$ & $223$ & $1,2,3,4,5,7,8$ & $J_{66}$ \\
     & $40$ & $239$ & $1,2,3,4,6,7,8$ & $J_{67}$ \\
 \hline
 $C$ & $41$ & $101$ & $1,3,6,7$ & $J_{68},J_{69},J_{70}$ \\
     & $42$ & $102$ & $2,3,6,7$ & $J_{71}$ \\
     & $43$ & $103$ & $1,2,3,6,7$ & $J_{72}$ \\
     & $44$ & $117$ & $1,3,5,6,7$ & $J_{73},J_{74},J_{75}$ \\
     & $45$ & $118$ & $2,3,5,6,7$ & $J_{76},J_{77},J_{78}$ \\
     & $46$ & $173$ & $1,3,4,6,8$ & $J_{79},J_{80}$ \\
     & $47$ & $181$ & $1,3,5,6,8$ & $J_{81}$ \\
     & $48$ & $119$ & $1,2,3,5,6,7$ & $J_{82},J_{83},J_{84}$ \\
     & $49$ & $175$ & $1,2,3,4,6,8$ & $J_{85}$ \\
     & $50$ & $183$ & $1,2,3,5,6,8$ & $J_{86}$ \\
     & $51$ & $189$ & $1,3,4,5,6,8$ & $J_{87},J_{88}$ \\
     & $52$ & $190$ & $2,3,4,5,6,8$ & $J_{89}$ \\
     & $53$ & $231$ & $1,2,3,6,7,8$ & $J_{90},J_{91},J_{92}$ \\
     & $54$ & $246$ & $2,3,5,6,7,8$ & $J_{93},J_{94}$ \\
     & $55$ & $191$ & $1,2,3,4,5,6,8$ & $J_{95},J_{96},J_{97},J_{98}$ \\
     & $56$ & $247$ & $1,2,3,5,6,7,8$ & $J_{99},J_{100},J_{101},J_{102}$ \\
     & $57$ & $253$ & $1,3,4,5,6,7,8$ & $J_{103}$ \\
     & $58$ & $255$ & $1,2,3,4,5,6,7,8$ & $J_{104}$ \\
 \hline
 $D$ & $59$ & $239$ & $1,2,3,4,6,7,8$ & $J_{105}$ \\
\hline
\end{tabular}
\end{center}
\caption{Overview of the set of master integrals (continued).
}
\label{table_master_integrals_BCD}
\end{table}


\section{The differential equation and differential forms}
\label{sect:differential_forms}

We write the differential equation for the master integrals as
\bq
\label{diff_eq_J}
 d J & = & A J,
\eq
with
\bq
 A & = & A_v \; dv + A_w \; dw.
\eq
The matrix-valued one-form $A$ satisfies the integrability condition
\bq
\label{integrability_condition}
 d A - A \wedge A & = & 0.
\eq
For the choice of master integrals as
in eq.~(\ref{def_master_integrals}),
the differential equation is in $\eps$-form \cite{Henn:2013pwa}
and we have
\bq
 A & = &
 \eps \sum\limits_{k=1}^{19} M_k \omega_k,
\eq
where the $M_k$'s are $105 \times 105$-matrices, whose entries are rational numbers.
The $\omega_k$'s are differential one-forms.
We have $19$ differential one-forms.
All are dlog-forms.
The differential one-forms are 
\bq
\label{def_differential_one_forms}
 \omega_{1}
 & = & 
 d \ln\left(w\right),
 \nonumber \\
 \omega_{2}
 & = & 
 d \ln\left(1-w\right),
 \nonumber \\
 \omega_{3}
 & = & 
 d \ln\left(-v\right),
 \nonumber \\
 \omega_{4}
 & = & 
 d \ln\left(1-v\right),
 \nonumber \\
 \omega_{5}
 & = & 
 d \ln\left(4-v\right),
 \nonumber \\
 \omega_{6}
 & = & 
 d \ln\left(w-v\right),
 \nonumber \\
 \omega_{7}
 & = & 
 d \ln\left(1-w-v\right),
 \nonumber \\
 \omega_{8}
 & = & 
 d \ln\left(1-w+v\right),
 \nonumber \\
 \omega_{9}
 & = & 
 d \ln\left(w^2+v\left(1-w\right)\right),
 \nonumber \\
 \omega_{10}
 & = & 
 d \ln\left(\left(1-w\right)^2+vw\right),
 \nonumber \\
 \omega_{11}
 & = & 
 d \ln\left(\left(1-w\right)^2+v\left(2-w\right)\right),
 \nonumber \\
 \omega_{12}
 & = & 
 d \ln\left(\lambda\left(v,w,1\right)\right),
 \nonumber \\
 \omega_{13}
 & = & 
 d \ln\left(\frac{2-v-r_1}{2-v+r_1}\right),
 \nonumber \\
 \omega_{14}
 & = & 
 d \ln\left(\frac{2w+v\left(1-w\right)-\left(1-w\right)r_1}{2w+v\left(1-w\right)+\left(1-w\right)r_1}\right),
 \nonumber \\
 \omega_{15}
 & = & 
 d \ln\left(\frac{2\left(1-w\right)+vw-wr_1}{2\left(1-w\right)+vw+wr_1}\right),
 \nonumber \\
 \omega_{16}
 & = & 
 d \ln\left(\frac{wv^2+\left(1-2w-w^2\right)v+2\left(1-w\right)^2-\left(1-w^2+vw\right)r_1}{wv^2+\left(1-2w-w^2\right)v+2\left(1-w\right)^2+\left(1-w^2+vw\right)r_1}\right),
 \nonumber \\
 \omega_{17}
 & = & 
 d \ln\left(\frac{1+w-v-r_2}{1+w-v+r_2}\right),
 \nonumber \\
 \omega_{18}
 & = & 
 d \ln\left(\frac{\left(1-w\right)^2-v\left(1+w\right)-\left(1-w\right)r_2}{\left(1-w\right)^2-v\left(1+w\right)+\left(1-w\right)r_2}\right),
 \nonumber \\
 \omega_{19}
 & = & 
 d \ln\left(\frac{-v\left(3-v+w\right)-r_1 r_2}{-v\left(3-v+w\right)+r_1 r_2}\right).
\eq
We write
\bq
 \omega_k & = & d\ln f_k.
\eq
The variable transformation from $(v,w)$ to $(x,y)$ rationalises all square roots.
The $f_k$'s are then rational functions in $x$ and $y$.
In detail we have
\begin{align}
 f_{1}
 & = 
 - p_1^{-1} p_5^{-1} p_6^{-1} p_8 p_9,
 &
 f_{2}
 & = 
 2 p_1^{-1} p_2 p_4 p_5^{-1} p_6^{-1} p_7,
 &
 f_{3}
 & = 
 p_1^{-1} p_2^{2},
 \nonumber \\
 f_{4}
 & = 
 p_1^{-1} p_{17},
 &
 f_{5}
 & = 
 p_1^{-1} p_3^{2},
 &
 f_{6}
 & = 
 - p_1^{-1} p_5^{-1} p_6^{-1} p_{18},
 \nonumber \\
 f_{7}
 & = 
 p_1^{-1} p_2 p_5^{-1} p_6^{-1} p_{16},
 &
 f_{8}
 & = 
 p_1^{-1} p_2 p_5^{-1} p_6^{-1} p_{12},
 &
 f_{9}
 & = 
 p_1^{-2} p_5^{-2} p_6^{-2} p_{19} p_{20},
 \nonumber \\
 f_{10}
 & = 
 p_1^{-2} p_2^{2} p_5^{-2} p_6^{-2} p_{14} p_{15},
 &
 f_{11}
 & = 
 p_1^{-2} p_2^{2} p_5^{-2} p_6^{-2} p_{10} p_{13},
 &
 f_{12}
 & = 
 p_1^{-2} p_2^{2} p_5^{-2} p_6^{-2} p_{11}^{2},
 \nonumber \\
 f_{13}
 & = 
 p_1^{2},
 &
 f_{14}
 & = 
 p_1^{-2} p_{19} p_{20}^{-1},
 &
 f_{15}
 & = 
 p_1^{-2} p_{14}^{-1} p_{15},
 \nonumber \\
 f_{16}
 & = 
 p_1^{-2} p_{10}^{-1} p_{13} p_{14}^{-1} p_{15},
 &
 f_{17}
 & = 
 - p_1 p_5^{-1} p_6 p_8 p_9^{-1},
 &
 f_{18}
 & = 
 - p_1^{-1} p_5^{3} p_6^{-3} p_8 p_9^{-1},
 \nonumber \\
 f_{19}
 & = 
 - 4 p_1 p_4^{2} p_7^{-2}.
\end{align}
The $p_j$'s are polynomials in $x$ and $y$ and given by
\bq
 p_1 & = & x,
 \nonumber \\
 p_2 & = & x - 1,
 \nonumber \\
 p_3 & = & x + 1,
 \nonumber \\
 p_4 & = & y,
 \nonumber \\
 p_5 & = & y - 1,
 \nonumber \\
 p_6 & = & y + 1,
 \nonumber \\
 p_7 & = & x y + x - y + 1,
 \nonumber \\
 p_8 & = & x y + x - 2 y,
 \nonumber \\
 p_9 & = & 2 x y - y + 1,
 \nonumber \\
 p_{10} & = & x y^2 + 2 x y - 2 y^2 + x + 2 y,
 \nonumber \\
 p_{11} & = & x y^2 + 2 x y + y^2 + x - 2 y + 1,
 \nonumber \\
 p_{12} & = & x y^2 + 2 x y - y^2 + x + 2 y - 1,
 \nonumber \\
 p_{13} & = & 2 x y^2 + 2 x y - y^2 + 2 y - 1,
 \nonumber \\
 p_{14} & = & 2 x y^2 + 2 x y - 3 y^2 + 2 y + 1,
 \nonumber \\
 p_{15} & = & 3 x y^2 + 2 x y - 2 y^2 - x + 2 y,
 \nonumber \\
 p_{16} & = & 3 x y^2 + 2 x y - 3 y^2 - x + 2 y + 1,
 \nonumber \\
 p_{17} & = & x^2 - x + 1,
 \nonumber \\
 p_{18} & = & x^2 y^2 + 2 x^2 y - 3 x y^2 + x^2 + y^2 - x - 2 y + 1,
 \nonumber \\
 p_{19} & = & x^2 y^2 + 2 x^2 y - 4 x y^2 + x^2 + 2 y^2 - 2 y,
 \nonumber \\
 p_{20} & = & 2 x^2 y^2 + 2 x^2 y - 4 x y^2 + y^2 - 2 y +1.
\eq
The polynomials $p_1$-$p_{16}$ are linear in $x$ and appeared previously
in the two-loop mixed QCD-electroweak corrections to $H\rightarrow b \bar{b}$ through a $Ht\bar{t}$-coupling \cite{Chaubey:2019lum}.
The polynomials $p_{17}$-$p_{20}$ are quadratic in $x$.

\section{Boundary values}
\label{sect:boundary}

In order to solve the differential equation we need boundary values.
As boundary point we take the point $(v,w)=(0,1)$, which corresponds in $(x,y)$-space to the line $x=1$.
The map from $(v,w)$-space to $(x,y)$-space blows up the point $(v,w)=(0,1)$.

Most master integrals vanish at the boundary point.
A few boundary values we have to compute explicitly. They are obtained from rather simple integrals.
We have
\begin{align}
 J_{1} & = C_{1} w^{-\eps},
 &
 J_{15} & = C_{15} \left(-v\right)^{-\eps},
 &
 J_{51} & = C_{51} \left(-v\right)^{-2\eps} w^{-\eps},
 &
 J_{60} & = C_{60} \left(-v\right)^{-2\eps},
 &
 J_{71} & = C_{71},
 &
\end{align}
with boundary values
\bq
\label{boundary_values_explicit}
 C_1 & = &
 e^{3 \gamma_E \eps} \left( \Gamma\left(1+\eps\right) \right)^3,
 \nonumber \\
 C_{15} & = &
 2 e^{3 \gamma_E \eps} \frac{\left( \Gamma\left(1+\eps\right) \right)^3 \left( \Gamma\left(1-\eps\right) \right)^2}{\Gamma\left(1-2\eps\right)},
 \nonumber \\
 C_{51} & = &
 -3 e^{3 \gamma_E \eps} \frac{\Gamma\left(1+\eps\right) \left( \Gamma\left(1-\eps\right) \right)^3\Gamma\left(1+2\eps\right)}{\Gamma\left(1-3\eps\right)},
 \nonumber \\
 C_{60} & = &
 4 e^{3 \gamma_E \eps} \frac{\left( \Gamma\left(1+\eps\right) \right)^3 \left( \Gamma\left(1-\eps\right) \right)^4}{\Gamma\left(1-2\eps\right)^2},
 \nonumber \\
 C_{71} & = &
 e^{3 \gamma_E \eps} \frac{\Gamma\left(1-\eps\right)\Gamma\left(1+\eps\right)\Gamma\left(1+2\eps\right)^2\Gamma\left(1+3\eps\right)}{\Gamma\left(1+4\eps\right)}.
\eq
Note that $J_{15}$, $J_{51}$ and $J_{60}$ have logarithmic singularities at $v=0$ and $C_{15}$, $C_{51}$, $C_{60}$ denote
``regularised'' boundary values: They are given by the finite values obtained after removing all powers of $\ln(-v)$.
The remaining non-zero boundary values are
\bq
 & &
 C_{14} \; = \; C_{1},
 \nonumber \\
 & &
 C_{16} \; = \;
 C_{59} \; = \; C_{15},
 \nonumber \\
 & &
 C_{52} \; = \; C_{51},
 \nonumber \\
 & &
 C_{61} \; = \; C_{60},
 \nonumber \\
 & &
 C_{6} \; = \;
 C_{8} \; = \; 
 C_{9} \; = \;
 C_{29} \; = \;
 C_{70} \; = \;
 C_{77} \; = \; \frac{1}{3} C_{71},
 \nonumber \\
 & &
 C_{56} \; = \; -\frac{2}{3} C_{71},
 \nonumber \\
 & &
 C_{30} \; = \; -\frac{1}{2} C_{1} + \frac{1}{2} C_{71},
 \nonumber \\
 & &
 C_{78} \; = \; \frac{1}{2} C_{1} - \frac{1}{2} C_{71},
 \nonumber \\
 & &
 C_{87} \; = \; -\frac{1}{4} C_{1} + \frac{1}{4} C_{71}.
\eq
All other boundary values are zero.

All non-trivial boundary values follow from
(i) the explicit computations of eq.~\ref{boundary_values_explicit},
(ii) a dedicated integral reduction for $s=0$ and $m_W^2=m_t^2$ and
(iii) the requirement that the master integrals $J_1$-$J_{105}$ are constant on the hypersurface $x=1$.
Note that this does not require that $A$ vanishes when restricted to $x=1$, it only requires that the vector $J$ is in the kernel of $A$ on
the hypersurface $x=1$.

\section{Analytical results}
\label{sect:analytical_results}

For the multiple polylogarithms \cite{Goncharov_no_note,Goncharov:2001,Borwein}
we use the notation $G(z_1,\dots,z_k;y)$, which is associated with the iterated integral representation.
One defines $G(0,...,0;y)$ with $k$ zeros to be
\bq
 G(0,...,0;y) & = & \frac{1}{k!} \left( \ln y \right)^k.
\eq
This includes the trivial case $G(;y)=1$.
Multiple polylogarithms are then defined recursively by
\bq
\label{Grecursive}
 G(z_1,z_2,...,z_k;y) & = & \int\limits_0^y \frac{dy_1}{y_1-z_1} G(z_2,...,z_k;y_1).
\eq
A discussion of the basic properties of multiple polylogarithms can be found in ref.~\cite{Weinzierl:2022eaz}.

We take the point $(v,w)=(0,1)$ in $(v,w)$-space as our boundary point.
In $(x,y)$-space this point corresponds to the hyperplane $x=1$. The master integrals are constant on this hyperplane.
We therefore have to integrate only along $x$ in $(x,y)$-space.
It is advantageous to change variables from $x$ to $x'=1-x$.
Doing so, the starting point of the integration  will be $x'=0$.
The integration gives multiple polylogarithms of the form
\bq
 G\left(l_1',...,l_k';x'\right),
\eq
where the letters $l_1'$, ..., $l_k'$ are from the
alphabet
\bq
 {\mathcal A}
 & = &
 \left\{ 
  0, 1, 2, x_7', x_8', x_9', x_{10}', x_{11}', x_{12}', x_{13}', x_{14}', x_{15}', x_{16}',
 \right. \nonumber \\
 & & \left.
  x_{17,a}', x_{17,b}',
  x_{18,a}', x_{18,b}',
  x_{19,a}', x_{19,b}',
  x_{20,a}', x_{20,b}'
 \right\}.
\eq
This is an alphabet with $21$ letters.
The non-trivial letters are given by
\begin{align}
 x_7' & = \frac{2}{1+y},
 &
 x_8' & = \frac{1-y}{1+y},
 &
 x_9' & = \frac{1+y}{2y},
 \nonumber \\
 x_{10}' & = \frac{1+4y-y^2}{\left(1+y\right)^2},
 &
 x_{11}' & = \frac{2 \left(1+y^2\right)}{\left(1+y\right)^2},
 &
 x_{12}' & = \frac{4y}{\left(1+y\right)^2},
 \nonumber \\
 x_{13}' & = - \frac{1-4y-y^2}{2 y \left(1+y\right)},
 &
 x_{14}' & = \frac{1+4y-y^2}{2 y \left(1+y\right)},
 &
 x_{15}' & = \frac{1-4y-y^2}{\left(1+y\right) \left(1-3y\right)},
 \nonumber \\
 x_{16}' & = - \frac{4 y}{\left(1+y\right) \left(1-3y\right)}
 & & & &
\end{align}
and
\begin{align}
 x_{17,a}'
 & =
 e^{\frac{\pi i}{3}},
 &
 x_{17,b}'
 & =
 e^{-\frac{\pi i}{3}},
 \nonumber \\
 x_{18,a}'
 & =
 \frac{1+4y-y^2+i\sqrt{\left(3+y^2\right)\left(1-5y^2\right)}}{2\left(1+y\right)^2},
 &
 x_{18,b}'
 & = 
 \frac{1+4y-y^2-i\sqrt{\left(3+y^2\right)\left(1-5y^2\right)}}{2\left(1+y\right)^2},
 \nonumber \\
 x_{19,a}'
 & = 
 \frac{1+2y-y^2+\sqrt{2y\left(1+y-y^2+y^3\right)}}{\left(1+y\right)^2},
 &
 x_{19,b}'
 & = 
 \frac{1+2y-y^2-\sqrt{2y\left(1+y-y^2+y^3\right)}}{\left(1+y\right)^2},
 \nonumber \\
 x_{20,a}'
 & =
 \frac{2y+i\sqrt{2y\left(1-y-y^2-y^3\right)}}{2y\left(1+y\right)},
 &
 x_{20,b}'
 & =
 \frac{2y-i\sqrt{2y\left(1-y-y^2-y^3\right)}}{2y\left(1+y\right)}.
\end{align}
The letters are the roots of the polynomials $p_1$-$p_{20}$ in $x'$ with $x$ substituted by $1-x'$.

For all master integrals we write
\bq
 J_k
 & = &
 \sum\limits_{j=0}^\infty \eps^j J_k^{(j)}.
\eq
The $\eps^j$-term $J_k^{(j)}$ is of uniform weight $j$.
To give an example, we have 
\bq
 J_{67}
 & = &
 2
 \left[
  G\left( 0, x_{8}', x_{9}'; x' \right)
  +G\left( 0, x_{9}', x_{8}'; x' \right)
 -G\left( x_{15}', x_{8}', 0; x' \right)
 -G\left( x_{14}', x_{9}', 0; x' \right)
 \right. \nonumber \\
 & & \left.
 -G\left( x_{14}', x_{8}', 0; x' \right)
 -G\left( x_{15}', x_{9}', 0; x' \right)
 +G\left( x_{14}', x_{8}', x_{9}'; x' \right)
 +G\left( x_{14}', x_{9}', x_{8}'; x' \right)
 \right. \nonumber \\
 & & \left.
 +G\left( x_{15}', x_{8}', x_{9}'; x' \right)
 +G\left( x_{15}', x_{9}', x_{8}'; x' \right)
 -G\left( 1, x_{9}', x_{8}'; x' \right)
 +G\left( x_{15}', x_{9}', 1; x' \right)
 \right. \nonumber \\
 & & \left.
 -G\left( x_{7}', x_{9}', 1; x' \right)
 +G\left( x_{7}', 1, x_{9}'; x' \right)
 -G\left( 0, x_{8}', 1; x' \right)
 -G\left( 1, x_{9}', 1; x' \right)
 \right. \nonumber \\
 & & \left.
 -G\left( 0, 1, x_{8}'; x' \right)
 -2\,G\left( x_{7}', 1, 0; x' \right)
 +G\left( 1, 0, 1; x' \right)
 -G\left( x_{14}', x_{8}', 1; x' \right)
 \right. \nonumber \\
 & & \left.
 +2\,G\left( x_{15}', 1, 0; x' \right)
 +2\,G\left( x_{7}', x_{8}', 0; x' \right)
 +2\,G\left( x_{7}', x_{9}', 0; x' \right)
 -G\left( x_{15}', x_{7}', 1; x' \right)
 \right. \nonumber \\
 & & \left.
 -G\left( x_{7}', x_{9}', x_{8}'; x' \right)
 -G\left( x_{15}', 1, x_{9}'; x' \right)
 -G\left( x_{7}', x_{8}', x_{9}'; x' \right)
 -G\left( x_{14}', x_{7}', x_{9}'; x' \right)
 \right. \nonumber \\
 & & \left.
 -G\left( x_{15}', x_{7}', x_{8}'; x' \right)
 -G\left( 1, x_{8}', x_{9}'; x' \right)
 +2\,G\left( x_{14}', x_{7}', 1; x' \right)
 +G\left( 1, x_{7}', 1; x' \right)
 \right. \nonumber \\
 & & \left.
 +G\left( 1, x_{7}', x_{8}'; x' \right)
 \right] \eps^3 
 + {\mathcal O}\left(\eps^4\right).
\eq
The analytical results for the master integrals are given up to the order $\eps^5$
in a supplementary electronic file attached to the arxiv version of this article.


\section{Numerical results}
\label{sect:numerical_results}

As a reference value we give numerical results for
\bq
 p^2
 & = &
 m_H^2.
\eq
Since $p^2>0$, we are not in the Euclidean region. Feynman's $i0$-prescription instructs us
to take a small imaginary part into account: $p^2 \rightarrow p^2 + i0$.
This selects the correct branches for the 
two square roots $\sqrt{-v(4-v)}$ and $\sqrt{\lambda(v,w,1)}$. With
\bq
 m_W
 \; = \;
 80.38 \; \mathrm{GeV},
 \;\;\;\;\;\;
 m_H
 \; = \;
 125.2 \; \mathrm{GeV},
 \;\;\;\;\;\;
 m_t
 \; = \;
 173.1 \; \mathrm{GeV}
\eq
we obtain for the
variables $x$ and $y$
\bq
 x \; = \; 0.7384 + 0.6743 i,
 & &
 y \; = \; 0.3987 i.
\eq
The values of the master integrals at this point are given 
to $8$ digits in tables~\ref{table_numerical_results_I} and~~\ref{table_numerical_results_II}.
They are easily computed to arbitrary precision by evaluating the multiple polylogarithms 
with the help of \verb|GiNaC| \cite{Bauer:2000cp,Vollinga:2004sn}.
In addition we verified the correctness of our results at several kinematic points with the help of the programs \verb|AMFlow| \cite{Liu:2022chg,Liu:2017jxz,Liu:2022mfb}
and
\verb|sector_decomposition| \cite{Bogner:2007cr,Binoth:2000ps}.
\begin{table}[!htbp]
\begin{center}
{\tiny
\begin{tabular}{|l|llllll|}
 \hline 
 & $\eps^0$ & $\eps^1$ & $\eps^2$ & $\eps^3$ & $\eps^4$ & $\eps^5$ \\
 \hline 
$J_{ 1}$ & $        1$ & $ 1.5342081$ & $ 3.6442984$ & $ 3.1853184$ & $ 5.1462966$ & $ 2.4686377$ \\ 
$J_{ 2}$ & $        0$ & $-1.4801083 i$ & $-2.4097495 i$ & $-5.6149488 i$ & $-5.2333001 i$ & $-8.0886313 i$ \\ 
$J_{ 3}$ & $        0$ & $ 1.4801083 i$ & $-0.22727279 i$ & $ 7.1496293 i$ & $-6.383137 i$ & $ 25.607456 i$ \\ 
$J_{ 4}$ & $        0$ & $        0$ & $-0.54768013$ & $-0.36476197$ & $-2.891637$ & $ 0.046729212$ \\ 
$J_{ 5}$ & $        0$ & $ 1.5342081$ & $-2.9983034$ & $ 19.075365$ & $-53.393746$ & $ 195.47467$ \\ 
$J_{ 6}$ & $ 0.33333333$ & $ 1.5342081$ & $-0.088235985$ & $ 15.450021$ & $-32.95737$ & $ 146.32479$ \\ 
$J_{ 7}$ & $        0$ & $-1.4585842 i$ & $ 1.294182 i$ & $-18.724204 i$ & $ 33.200239 i$ & $-173.64768 i$ \\ 
$J_{ 8}$ & $ 0.33333333$ & $        0$ & $ 3.7496407$ & $-3.6797304$ & $ 30.246157$ & $-60.506131$ \\ 
$J_{ 9}$ & $ 0.33333333$ & $ 1.5342081$ & $ 1.3128219$ & $ 16.578415$ & $-17.750719$ & $ 142.4298$ \\ 
$J_{ 10}$ & $        0$ & $        0$ & $-2.2406312$ & $ 1.0647958$ & $-26.559377$ & $ 40.135978$ \\ 
$J_{ 11}$ & $        0$ & $-1.5342081$ & $ 0.91070302$ & $-7.2586694$ & $ 10.224944$ & $-28.168258$ \\ 
$J_{ 12}$ & $        0$ & $ 1.4585842 i$ & $ 0.050868134 i$ & $ 7.6533633 i$ & $-4.6991476 i$ & $ 27.583113 i$ \\ 
$J_{ 13}$ & $        0$ & $ 1.5342081$ & $-1.7502763$ & $ 7.0599975$ & $-14.845047$ & $ 30.115459$ \\ 
$J_{ 14}$ & $        1$ & $ 1.5342081$ & $ 5.8849296$ & $ 4.6861424$ & $ 17.388435$ & $ 2.5361954$ \\ 
$J_{ 15}$ & $        2$ & $ 1.295828 +  6.2831853 i$ & $-7.8048778 +  4.0709638 i$ & $-12.450557 -3.8488956 i$ & $-11.660945 -25.721646 i$ & $ 10.668736 -35.695423 i$ \\ 
$J_{ 16}$ & $        2$ & $ 4.3642443 +  6.2831853 i$ & $-3.4630132 +  13.710678 i$ & $-21.696071 +  9.7914742 i$ & $-38.706617 -23.053895 i$ & $-26.131312 -75.786891 i$ \\ 
$J_{ 17}$ & $        0$ & $        0$ & $ 2.1907205$ & $ 3.772359$ & $ 8.6571337$ & $ 8.5453519$ \\ 
$J_{ 18}$ & $        0$ & $-1.4801083 i$ & $ 0.5445262 i$ & $-15.995422 i$ & $ 23.86858 i$ & $-132.81753 i$ \\ 
$J_{ 19}$ & $        0$ & $        0$ & $-5.2250698 i$ & $ 10.295242 i$ & $-65.298537 i$ & $ 183.26131 i$ \\ 
$J_{ 20}$ & $        0$ & $        0$ & $ 2.2707942 i$ & $-1.1347528 i$ & $ 10.629046 i$ & $-14.132023 i$ \\ 
$J_{ 21}$ & $        0$ & $ 1.4801083 i$ & $-5.4523426 i$ & $ 28.891559 i$ & $-88.331183 i$ & $ 331.78433 i$ \\ 
$J_{ 22}$ & $        0$ & $-1.5342081$ & $ 3.2901965$ & $-18.78899$ & $ 54.646442$ & $-195.09957$ \\ 
$J_{ 23}$ & $        0$ & $ 1.5342081$ & $-2.1449256$ & $ 21.075678$ & $-46.255399$ & $ 207.64266$ \\ 
$J_{ 24}$ & $        0$ & $        0$ & $ 2.1588625$ & $ 0.27796832$ & $ 11.346261$ & $-5.8909184$ \\ 
$J_{ 25}$ & $        0$ & $        0$ & $-2.2707942 i$ & $ 2.3774122 i$ & $-10.218327 i$ & $ 21.004436 i$ \\ 
$J_{ 26}$ & $        0$ & $-1.4801083 i$ & $-2.4097495 i$ & $-8.9313256 i$ & $-7.7660295 i$ & $-26.43436 i$ \\ 
$J_{ 27}$ & $        0$ & $-1.4585842 i$ & $ 1.5911765 i$ & $-16.821522 i$ & $ 38.929474 i$ & $-158.55603 i$ \\ 
$J_{ 28}$ & $        0$ & $-1.4801083 i$ & $ 5.4523426 i$ & $-27.832513 i$ & $ 90.872416 i$ & $-322.88758 i$ \\ 
$J_{ 29}$ & $ 0.33333333$ & $        0$ & $ 4.2973208$ & $-7.8296127$ & $ 39.354989$ & $-110.36825$ \\ 
$J_{ 30}$ & $        0$ & $        0$ & $        0$ & $ 3.9327313$ & $-9.7902493$ & $ 50.090814$ \\ 
$J_{ 31}$ & $        0$ & $-1.4801083 i$ & $ 2.8153204 i$ & $-9.1045495 i$ & $ 18.752518 i$ & $-45.336109 i$ \\ 
$J_{ 32}$ & $        0$ & $-2.9602166 i$ & $ 9.2997946 -2.1958763 i$ & $ 6.898549 +  11.35639 i$ & $-5.0820522 +  19.501961 i$ & $ -38.5719 +  19.050034 i$ \\ 
$J_{ 33}$ & $        0$ & $ 2.9602166 i$ & $-9.2997946 -3.0781682 i$ & $ 9.6703505 -3.6126738 i$ & $-19.245548 -12.824796 i$ & $ 72.104463 +  0.12405401 i$ \\ 
$J_{ 34}$ & $        0$ & $        0$ & $-1.0953603$ & $ 0.24128744 -3.4411758 i$ & $ 1.6402683 +  0.75802684 i$ & $ 6.1054746 -6.1679597 i$ \\ 
$J_{ 35}$ & $        0$ & $-2.9602166 i$ & $ 9.2997946 -6.7374647 i$ & $ 21.16637 +  4.5035873 i$ & $ 16.446661 +  32.559044 i$ & $-32.652689 +  60.330416 i$ \\ 
$J_{ 36}$ & $        0$ & $-1.5342081$ & $-2.6034827 -9.6397141 i$ & $  26.9974 -0.52358996 i$ & $-27.747558 -12.210078 i$ & $ 223.69832 +  60.357017 i$ \\ 
$J_{ 37}$ & $        0$ & $        0$ & $        0$ & $        0$ & $ 2.4095606$ & $-7.4050612$ \\ 
$J_{ 38}$ & $        0$ & $        0$ & $        0$ & $-0.80345343$ & $ 2.3774716$ & $-10.59662$ \\ 
$J_{ 39}$ & $        0$ & $        0$ & $ 2.1907205$ & $-3.9613101$ & $ 13.09607$ & $-26.512146$ \\ 
$J_{ 40}$ & $        0$ & $ 2.9602166 i$ & $-9.2997946 -3.712675 i$ & $ 14.988491 -6.3598329 i$ & $-8.0684303  -1.80014 i$ & $ 80.59546 +  18.841397 i$ \\ 
$J_{ 41}$ & $        0$ & $        0$ & $        0$ & $-1.2199963 -1.2715067 i$ & $ 5.3143264 -0.86415471 i$ & $-13.01594 -2.7052663 i$ \\ 
$J_{ 42}$ & $        0$ & $        0$ & $        0$ & $        0$ & $ 4.2051087 +  3.3549988 i$ & $-14.946099 +  3.7606676 i$ \\ 
$J_{ 43}$ & $        0$ & $        0$ & $        0$ & $        0$ & $ 2.6414875 +  2.2409565 i$ & $-11.227012 +  0.75555211 i$ \\ 
$J_{ 44}$ & $        0$ & $        0$ & $        0$ & $-1.2199963 -1.2715067 i$ & $  4.98217 -0.70082991 i$ & $-11.56359 -2.4458449 i$ \\ 
$J_{ 45}$ & $        0$ & $        0$ & $ 4.381441$ & $ 3.6614725 +  13.764703 i$ & $-16.48045 +  11.502855 i$ & $-30.424976 -6.4908014 i$ \\ 
$J_{ 46}$ & $        0$ & $        0$ & $        0$ & $-1.2199963 -1.2715067 i$ & $ 3.4643984 -1.4351478 i$ & $-2.3605461 +  0.17411836 i$ \\ 
$J_{ 47}$ & $        0$ & $        0$ & $        0$ & $        0$ & $        0$ & $ 2.7105126 +  2.3055744 i$ \\ 
$J_{ 48}$ & $        0$ & $        0$ & $        0$ & $        0$ & $ 0.65371112 +  1.0931833 i$ & $-1.0109578 +  3.0469453 i$ \\ 
$J_{ 49}$ & $        0$ & $        0$ & $        0$ & $        0$ & $ 1.6251574 +  0.75070903 i$ & $-1.068945 +  3.6561297 i$ \\ 
$J_{ 50}$ & $        0$ & $        0$ & $        0$ & $        0$ & $ 0.94098377 -0.90286329 i$ & $ 1.1504284 +  2.4790799 i$ \\ 
$J_{ 51}$ & $       -3$ & $-3.887484 -18.849556 i$ & $ 59.166272 -24.425783 i$ & $ 118.51311 +  123.70244 i$ & $-92.180991 +  423.20939 i$ & $-842.91965 +  395.8266 i$ \\ 
$J_{ 52}$ & $       -3$ & $-8.4901085 -18.849556 i$ & $ 49.671371 -53.344925 i$ & $ 202.90571 +  64.044212 i$ & $ 156.24311 +  572.90309 i$ & $-810.36632 +  1171.6505 i$ \\ 
$J_{ 53}$ & $        0$ & $ 4.4403248 i$ & $-27.899384 +  6.1707632 i$ & $-38.772048 -87.008884 i$ & $ 179.55177 -183.60241 i$ & $ 643.38826 +  119.50921 i$ \\ 
$J_{ 54}$ & $        0$ & $-2.9171683 i$ & $ 9.1645545 -1.9918105 i$ & $ 24.127791 +  5.3212259 i$ & $-21.804448 +  120.95424 i$ & $-196.92433 -102.13413 i$ \\ 
$J_{ 55}$ & $        0$ & $-1.5342081$ & $-8.2052687 -19.279428 i$ & $ 79.333963 -36.244239 i$ & $ 118.23648 +  121.5975 i$ & $ 181.95398 +  573.21464 i$ \\ 
$J_{ 56}$ & $-0.66666667$ & $-1.5342081$ & $-7.3030938$ & $-22.007405 -14.078301 i$ & $-7.7964625 -73.29467 i$ & $ 88.637395 -153.09033 i$ \\ 
$J_{ 57}$ & $        0$ & $ 2.9171683 i$ & $-9.1645545 +  1.9918105 i$ & $-6.2574572 -3.2092524 i$ & $-20.068012 -19.446051 i$ & $ 40.505362 -29.233598 i$ \\ 
$J_{ 58}$ & $        0$ & $ 3.0684163$ & $-1.5124828 +  9.6397141 i$ & $-7.6933829 -4.7516047 i$ & $-18.562293 +  7.543913 i$ & $-4.8730848 -73.947317 i$ \\ 
$J_{ 59}$ & $        2$ & $ 4.3642443 +  6.2831853 i$ & $-1.2223821 +  13.710678 i$ & $-18.743511 +  16.830625 i$ & $-39.764537 -13.778153 i$ & $-42.529113 -55.952566 i$ \\ 
$J_{ 60}$ & $        4$ & $ 5.1833121 +  25.132741 i$ & $-78.888363 +  32.56771 i$ & $-129.16811 -164.93658 i$ & $ 199.25544 -383.01328 i$ & $ 779.6171 -48.063607 i$ \\ 
$J_{ 61}$ & $        4$ & $ 11.320145 +  25.132741 i$ & $-66.228494 +  71.126566 i$ & $-241.69158 -85.392283 i$ & $-87.715765 -582.60488 i$ & $ 887.29829 -804.39581 i$ \\ 
$J_{ 62}$ & $        0$ & $        0$ & $        0$ & $-1.2199963 -1.2715067 i$ & $ 5.9914141 -3.4779304 i$ & $-0.18199463 +  1.5542242 i$ \\ 
$J_{ 63}$ & $        0$ & $ 2.9602166 i$ & $-9.2997946 -3.712675 i$ & $-16.248795 -0.64764184 i$ & $ 40.454671 -155.25155 i$ & $ 168.63449 +  314.27487 i$ \\ 
$J_{ 64}$ & $        0$ & $-2.9602166 i$ & $ 9.2997946 +  3.712675 i$ & $-11.663712 +  4.2951778 i$ & $ 17.101399 +  9.4430398 i$ & $-68.03826 -7.8262762 i$ \\ 
$J_{ 65}$ & $        0$ & $-5.9204331 i$ & $ 37.199178 -8.2276842 i$ & $ 51.696065 +  116.01185 i$ & $-239.40236 +  202.10303 i$ & $-589.55784 -276.35688 i$ \\ 
$J_{ 66}$ & $        0$ & $        0$ & $        0$ & $        0$ & $        0$ & $ 2.6632592 +  5.3571394 i$ \\ 
$J_{ 67}$ & $        0$ & $        0$ & $        0$ & $ 1.2199963 +  1.2715067 i$ & $-6.6685018 +  6.0917061 i$ & $-14.752046 -15.744008 i$ \\ 
 \hline 
\end{tabular}
}
\end{center}
\caption{
Numerical results for the first six terms of the $\eps$-expansion of the master integrals $J_{1}$-$J_{67}$ 
at the kinematic point $p^2=m_H^2$.
}
\label{table_numerical_results_I}
\end{table}
\begin{table}[!htbp]
\begin{center}
{\tiny
\begin{tabular}{|l|llllll|}
 \hline 
 & $\eps^0$ & $\eps^1$ & $\eps^2$ & $\eps^3$ & $\eps^4$ & $\eps^5$ \\
 \hline 
$J_{ 68}$ & $        0$ & $-1.4801083 i$ & $ 5.1350892 i$ & $-25.837256 i$ & $ 86.938604 i$ & $-308.88849 i$ \\ 
$J_{ 69}$ & $        0$ & $        0$ & $ 1.0953603$ & $-3.0231166$ & $ 16.571142$ & $-51.663494$ \\ 
$J_{ 70}$ & $ 0.33333333$ & $        0$ & $ 1.3701472$ & $ 1.8283464$ & $-4.959024$ & $ 38.837723$ \\ 
$J_{ 71}$ & $        1$ & $        0$ & $ 2.4674011$ & $ 8.4143983$ & $-35.10786$ & $ 169.45714$ \\ 
$J_{ 72}$ & $        0$ & $-1.4801083 i$ & $ 2.498067 i$ & $-15.01875 i$ & $ 37.560716 i$ & $-132.37051 i$ \\ 
$J_{ 73}$ & $        0$ & $-1.4585842 i$ & $ 1.5911765 i$ & $-15.779761 i$ & $ 37.376156 i$ & $-152.62396 i$ \\ 
$J_{ 74}$ & $        0$ & $ 1.5342081$ & $-2.1449256$ & $ 18.664175$ & $-47.494038$ & $ 189.76691$ \\ 
$J_{ 75}$ & $        0$ & $-1.5342081$ & $-0.2345679$ & $-8.9971424$ & $ 3.8407769$ & $-34.623785$ \\ 
$J_{ 76}$ & $        0$ & $-1.4585842 i$ & $ 1.294182 i$ & $-16.104568 i$ & $ 36.151551 i$ & $-153.04501 i$ \\ 
$J_{ 77}$ & $ 0.33333333$ & $ 1.5342081$ & $ 0.62627943$ & $ 18.031643$ & $-37.794484$ & $ 187.99495$ \\ 
$J_{ 78}$ & $        0$ & $        0$ & $        0$ & $ 1.5692999$ & $-6.5217998$ & $ 25.959061$ \\ 
$J_{ 79}$ & $        0$ & $ 2.9602166 i$ & $-9.2997946 -3.0781682 i$ & $-14.67018 -1.6806018 i$ & $ 50.404394 -150.6488 i$ & $ 194.75027 +  331.14331 i$ \\ 
$J_{ 80}$ & $        0$ & $        0$ & $-0.54768013$ & $-0.41446154 -1.7205879 i$ & $-2.3612249 -7.5128487 i$ & $ 29.563404 -11.514285 i$ \\ 
$J_{ 81}$ & $        0$ & $ 1.5342081$ & $-2.9983034$ & $ 17.247662$ & $-52.439144$ & $ 184.22811$ \\ 
$J_{ 82}$ & $        0$ & $ 1.4801083 i$ & $-5.4523426 i$ & $ 23.141893 i$ & $-87.484869 i$ & $ 292.60701 i$ \\ 
$J_{ 83}$ & $        0$ & $        0$ & $        0$ & $-0.80345343$ & $ 2.4731997$ & $-10.583527$ \\ 
$J_{ 84}$ & $        0$ & $        0$ & $        0$ & $        0$ & $ 1.6505684$ & $-6.5723197$ \\ 
$J_{ 85}$ & $        0$ & $        0$ & $        0$ & $-0.9800827 -0.86029394 i$ & $ 4.9546534 -1.9953368 i$ & $-4.239423 +  1.9524939 i$ \\ 
$J_{ 86}$ & $        0$ & $        0$ & $-0.31725341 i$ & $-1.4970783 i$ & $ 1.498074 i$ & $-9.4006924 i$ \\ 
$J_{ 87}$ & $        0$ & $        0$ & $-0.13692003$ & $ 2.3004984 -0.43014697 i$ & $-6.0933617 -0.63742354 i$ & $ 39.061373 +  3.0381966 i$ \\ 
$J_{ 88}$ & $        0$ & $        0$ & $        0$ & $-1.2199963 -1.2715067 i$ & $ 5.1907304 -3.6242119 i$ & $ 2.1054694 +  0.72653192 i$ \\ 
$J_{ 89}$ & $        0$ & $        0$ & $        0$ & $-1.3411296 -1.3187987 i$ & $ 5.076922 -3.9113363 i$ & $ 2.0464347 -0.00022367332 i$ \\ 
$J_{ 90}$ & $        0$ & $        0$ & $        0$ & $-0.9800827 -0.86029394 i$ & $ 4.8178105 +  0.24314247 i$ & $-14.201072 -1.4422657 i$ \\ 
$J_{ 91}$ & $        0$ & $        0$ & $        0$ & $        0$ & $ 2.4985099 +  1.7074518 i$ & $-12.225556 -0.48055535 i$ \\ 
$J_{ 92}$ & $        0$ & $ 1.4801083 i$ & $-9.2997946 +  2.0569211 i$ & $ 2.2589826 -23.245834 i$ & $-5.6384268 +  23.974917 i$ & $ 56.355336 -142.57165 i$ \\ 
$J_{ 93}$ & $        0$ & $        0$ & $        0$ & $        0$ & $ 2.7500878 +  2.2730554 i$ & $-11.219892 +  0.89738075 i$ \\ 
$J_{ 94}$ & $        0$ & $        0$ & $        0$ & $-1.3411296 -1.3187987 i$ & $ 4.9222387 -0.91431241 i$ & $-11.726836 -2.8899212 i$ \\ 
$J_{ 95}$ & $        0$ & $        0$ & $        0$ & $        0$ & $        0$ & $ 1.8404286 +  0.14449436 i$ \\ 
$J_{ 96}$ & $        0$ & $        0$ & $        0$ & $        0$ & $ 0.21789937 +  0.042131655 i$ & $ 0.50386273 +  0.33655445 i$ \\ 
$J_{ 97}$ & $        0$ & $        0$ & $        0$ & $-0.23991356 -0.41121273 i$ & $ 0.85908359 -1.5044267 i$ & $ 3.8455066 -0.56878995 i$ \\ 
$J_{ 98}$ & $        0$ & $        0$ & $        0$ & $        0$ & $ 0.78680569 -5.3224605 i$ & $ 4.5796027 +  13.058166 i$ \\ 
$J_{ 99}$ & $        0$ & $        0$ & $        0$ & $        0$ & $        0$ & $ 2.8101582 +  2.3220016 i$ \\ 
$J_{ 100}$ & $        0$ & $        0$ & $        0$ & $        0$ & $ 0.39649658 +  0.58554103 i$ & $-0.70924343 +  1.3293236 i$ \\ 
$J_{ 101}$ & $        0$ & $        0$ & $        0$ & $-0.36104693 -0.45850473 i$ & $ 0.35197787 -1.3297874 i$ & $ 0.96759852 -1.7438646 i$ \\ 
$J_{ 102}$ & $        0$ & $        0$ & $        0$ & $        0$ & $ 0.96352362 -0.98327656 i$ & $ 2.3321043 +  1.5761209 i$ \\ 
$J_{ 103}$ & $        0$ & $        0$ & $        0$ & $        0$ & $        0$ & $ 1.9661517 +  5.3108372 i$ \\ 
$J_{ 104}$ & $        0$ & $        0$ & $        0$ & $        0$ & $        0$ & $ 0.022784811 +  0.43623285 i$ \\ 
$J_{ 105}$ & $        0$ & $        0$ & $        0$ & $        0$ & $ 0.10737705 +  0.15759037 i$ & $-0.197588 +  0.3448854 i$ \\ 
 \hline 
\end{tabular}
}
\end{center}
\caption{
Numerical results for the first six terms of the $\eps$-expansion of the master integrals $J_{68}$-$J_{105}$ 
at the kinematic point $p^2=m_H^2$.
}
\label{table_numerical_results_II}
\end{table}

\section{Conclusions}
\label{sect:conclusions}

In this paper we considered all master integrals with internal top-and $W$-propagators
contributing to the three-loop Higgs self-energy diagrams of order ${\mathcal O}(\alpha^2 \alpha_s)$.
We obtained analytic results for all master integrals, keeping the full dependence on $p^2$, $m_W^2$ and $m_t^2$.
We presented a basis of master integrals of uniform weight.
In this basis the associated differential equation is $\eps$-factorised.
We set up three independent systems of differential equations, such that for each system all occurring square roots
can be rationalised simultaneously.
This allows us to express all master integrals to any order in the dimensional regularisation parameter $\eps$ in terms of multiple polylogarithms.
Our results are a building block for precision Higgs physics.

\subsection*{Acknowledgements}

We thank Johann Usovitsch for support with the computer program {\tt Kira}.
E.C. receives funding from the European Union’s Horizon 2020 research and innovation programme 
``High precision multi-jet dynamics at the LHC'' (consolidator grant agreement No. 772099).
S.W. would like to thank the Munich Institute for Astro-, Particle and BioPhysics (MIAPbP)
for hospitality, where this work was finalised.


\begin{appendix}

\section{The additional topologies}
\label{sect:additional_topologies}

\begin{figure}
\begin{center}
\includegraphics[scale=1.0]{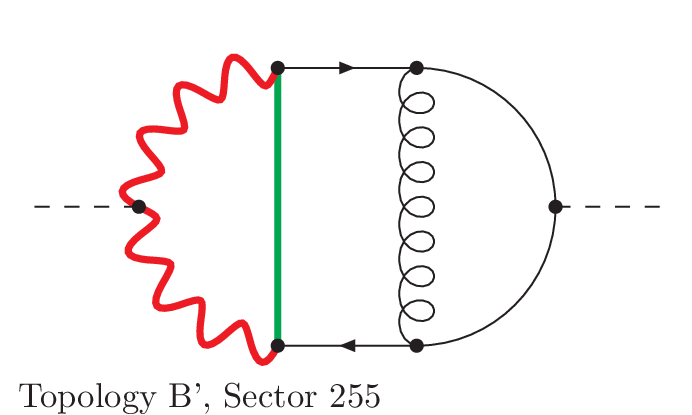}
\hspace*{10mm}
\includegraphics[scale=1.0]{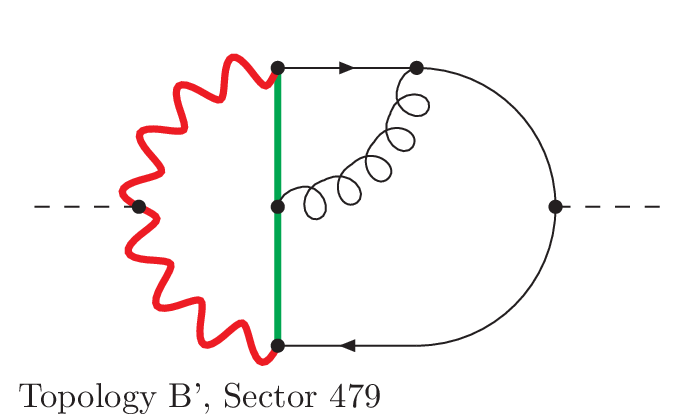}
\\
\includegraphics[scale=1.0]{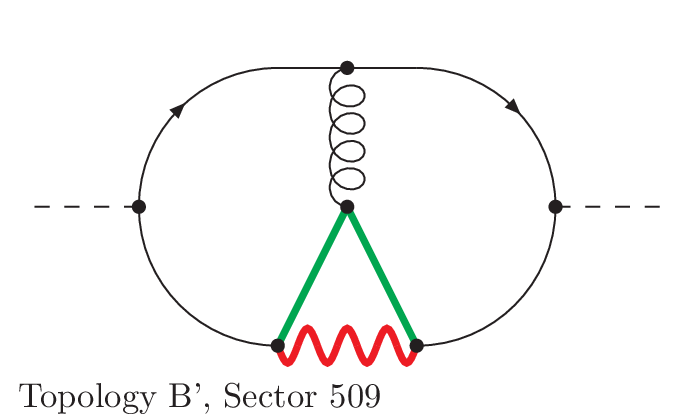}
\end{center}
\caption{
Examples of additional three-loop Higgs self-energy diagrams containing a top propagator and a $W$-propagator
and not being proportional to the product of Yukawa couplings $y_b y_t$.
The internal masses of the propagators are encoded by the colour of the propagators: 
massless (black), $m_t$ (green), $m_W$ (red).
}
\label{fig_extra_diagrams}
\end{figure}

\begin{figure}
\begin{center}
\includegraphics[scale=1.0]{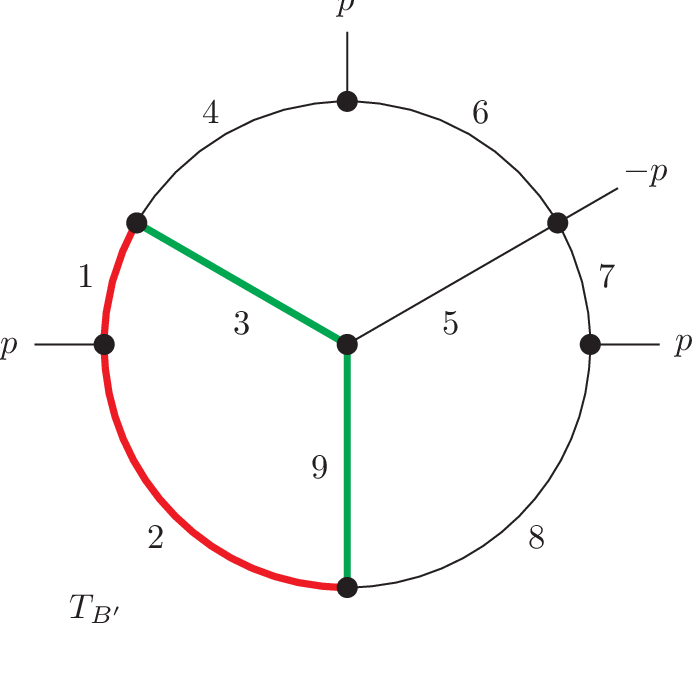}
\end{center}
\caption{
The topology $B'$. This topology is related by the exchange $m_W \leftrightarrow m_t$ to topology $B$.
}
\label{fig_auxiliary_topology_Bp}
\end{figure}

In this appendix we consider the additional topologies containing a top propagator and a $W$-propagator
and not being proportional to the product of Yukawa couplings $y_b y_t$.
There are three top sector diagrams which we have to consider.
These are shown in fig.~\ref{fig_extra_diagrams}.
The diagrams shown in fig.~\ref{fig_extra_diagrams} have all eight propagators and define the top sectors.
Simpler diagrams, obtained from the ones shown in fig.~\ref{fig_extra_diagrams} by pinching a subset of the propagators,
are not shown.
The three top sectors can be obtained from one auxiliary graph with nine propagators, shown in fig.~\ref{fig_auxiliary_topology_Bp}.
The auxiliary graph $B'$ is obtained from the auxiliary graph $B$ through the exchange $m_W \leftrightarrow m_t$.
The three diagrams in fig.~\ref{fig_extra_diagrams} correspond to the sectors $255$, $479$ and $509$ of the auxiliary graph $B'$.
The auxiliary graph $B'$ introduces additional square roots
\bq
 r_3
 \; = \;
 \sqrt{-v\left(4w-v\right)},
 & &
 r_4
 \; = \;
 \sqrt{-v\left(-4-v\right)}.
\eq
For the sectors $B_{255}'$ and $B_{479}'$ we encounter the set of square roots $(r_1,r_2,r_3)$, 
while for the sector $B_{509}'$ we encounter the set of square roots $(r_1,r_2,r_4)$.
We do not know how to rationalise all four square roots simultaneously.
But this is not needed: It is sufficient being able to rationalise the set of square roots $(r_1,r_2,r_3)$
and independently the set of square roots $(r_1,r_2,r_4)$.
This can be done.
We are therefore led to two system of differential equations, one for the sectors $B_{255}'$ and $B_{479}'$ 
with $61$ master integrals $K_{1}$-$K_{61}$
and another system for the sector $B_{509}'$ with $43$ master integrals $L_{1}$-$L_{43}$.
In both cases we are able to transform the differential equation into an $\eps$-factorised form
\bq
\label{diff_eq_K_L}
 d K & = & B K,
 \;\;\;\;\;\;
 B \; = \; \eps \sum\limits_{k=1}^{28} M_k' \omega_k,
 \nonumber \\
 d L & = & C L,
 \;\;\;\;\;\;
 C \; = \; \eps \sum\limits_{k=1}^{28} M_k'' \omega_k.
\eq
The entries of the $61 \times 61$ matrices $M_k'$ and of the $43 \times 43$ matrices $M_k''$
are rational numbers.
The differential one-forms 
$\omega_1$-$\omega_{19}$ are given in eq.~(\ref{def_differential_one_forms}),
the additional differential one-forms are given by
\bq
 \omega_{20}
 & = & 
 d \ln\left(4w-v\right),
 \nonumber \\
 \omega_{21}
 & = & 
 d \ln\left(\left(1-w\right)^2+v\right),
 \nonumber \\
 \omega_{22}
 & = & 
 d \ln\left(\frac{2w-v-r_3}{2w-v+r_3}\right),
 \nonumber \\
 \omega_{23}
 & = & 
 d \ln\left(\frac{2w\left(1-w\right)-v-r_3}{2w\left(1-w\right)-v+r_3}\right),
 \nonumber \\
 \omega_{24}
 & = & 
 d \ln\left(\frac{-v\left(1-v+3w\right)-r_2r_3}{-v\left(1-v+3w\right)+r_2r_3}\right),
 \nonumber \\
 \omega_{25}
 & = & 
 d \ln\left(4+v\right),
 \nonumber \\
 \omega_{26}
 & = & 
 d \ln\left(\left(1-w\right)^2-vw\right),
 \nonumber \\
 \omega_{27}
 & = & 
 d \ln\left(\frac{2+v-r_4}{2+v+r_4}\right),
 \nonumber \\
 \omega_{28}
 & = & 
 d \ln\left(\frac{-v\left(1+w\right)-\left(1-w\right)r_4}{-v\left(1+w\right)+\left(1-w\right)r_4}\right).
\eq
Not all differential one-forms do occur in the two differential equation in 
eq.~(\ref{diff_eq_K_L}).
The occurring ones are
\bq
 K: & &
 \left\{
  \omega_{1}, \omega_{2}, \omega_{3},
  \omega_{5}, \omega_{6},
  \omega_{8},
  \omega_{12}, \omega_{13},
  \omega_{17}, \omega_{18}, \omega_{19},
  \omega_{20}, \omega_{21}, \omega_{22}, \omega_{23}, \omega_{24}
 \right\},
 \nonumber \\
 L: & &
 \left\{
  \omega_{1}, \omega_{2}, \omega_{3},
  \omega_{5}, \omega_{6},
  \omega_{8},
  \omega_{12}, \omega_{13},
  \omega_{17}, \omega_{18}, \omega_{19},
  \omega_{25}, \omega_{26}, \omega_{27}, \omega_{28}
 \right\}.
\eq
We remark that the three systems of master integrals $(A_{255},B_{255},C_{255},D_{255})$,
$(B_{255}',B_{479}')$ and $(B_{509}')$ are redundant: The master integrals of $B_{255}'$
are related through the exchange $m_W \leftrightarrow m_t$ to the ones of $B_{255}$.
The master integrals of $B_{255}'$ are $K_{1}$-$K_{32}$, hence only $K_{33}$-$K_{61}$ are ``new''.
The same situation appears in the system $B_{509}'$: The master integrals $L_{1}$-$L_{39}$ appear already in the 
system $(B_{255}',B_{479}')$, hence only $L_{40}$-$L_{43}$ are ``new''.
This redundancy is the price to pay for having individual systems which can be solved independently with a single
rationalisation.

\subsection{The sectors $B_{255}'$ and $B_{479}'$}

The sector $B_{479}'$ involves the square roots $(r_1,r_2,r_3)$.
(The sector $B_{255}'$ alone involves only the square roots $(r_2,r_3)$.)
The square roots $(r_1,r_2,r_3)$  are rationalised by the variables $(\hat{x},y)$.
The variable $\hat{x}$ is defined by
\bq
 x & = &
 \frac{\left(1-y\right)}{\left(1+y\right)}
 \frac{\left[\left(1+y\right)\left(1-4\hat{x}\right)+\left(3+11y\right)\hat{x}^2\right]}{\left[1-y-\left(1+7y\right)\hat{x}^2\right]}, 
 \nonumber \\
 \hat{x}
 & = &
 \frac{\left(1-y^2\right)}{\left(1-x\right)} 
 \frac{\left[2\left(1-x\right)-xr_3\right]}{\left[3+x+8y-11y^2+8xy+7xy^2\right]}.
\eq
This rationalisation is easily obtained with the help of the computer program \verb|RationalizeRoots| \cite{Besier:2018jen,Besier:2019aaa}.
The value $x=1$ corresponds to $\hat{x}=0$.
The root $r_3$ is given in the variables $(\hat{x},y)$ by
\bq
 r_3
 & = &
 \frac{8 \hat{x}\left[1-y^2-\left(1+4y-y^2\right)\hat{x}\right]\left[1-y^2-2\left(1+4y-y^2\right)\hat{x}+\left(1+y\right)\left(1+7y\right)\hat{x}^2\right]}{\left(1-y^2\right)\left[1-y-\left(1+7y\right)\hat{x}^2\right]\left[\left(1+y\right)\left(1-4\hat{x}\right)+\left(3+11y\right)\hat{x}^2\right]}.
\eq 
With this rationalisation the differential one-forms translate to
\bq
 d\ln\left(\hat{p}_{1}\right), \dots, d\ln\left(\hat{p}_{20}\right),
\eq
where $\hat{p}_{1}$-$\hat{p}_{20}$ are polynomials in $(\hat{x},y)$.
We note that the differential one-form $\omega_6=d\ln(w-v)$ leads 
to a quartic polynomial in $\hat{x}$:
\bq
 \hat{p}_{20}
 & = &
 1-4\,\hat{x}-2\,{y}^{2}-16\,y\hat{x}+18\,{\hat{x}}^{2}+8\,{y}^{2}\hat{x}+48\,{\hat{x}}^{2}y-28\,{\hat{x}}^{3}+{y}^{4}+16\,{y}^{3}\hat{x}
 +44\,{\hat{x}}^{2}{y}^{2}-144\,{\hat{x}}^{3}y
 \nonumber \\
 & &
 +13\,{\hat{x}}^{4}-4\,{y}^{4}\hat{x}-48\,{\hat{x}}^{2}{y}^{3}-104\,{y}^{2}{\hat{x}}^{3}+112\,{\hat{x}}^{4}y+2\,{y}^{4}{\hat{x}}^{2}
 +16\,{y}^{3}{\hat{x}}^{3}+246\,{\hat{x}}^{4}{y}^{2}+4\,{y}^{4}{\hat{x}}^{3}
 \nonumber \\
 & &
 +16\,{y}^{3}{\hat{x}}^{4}+61\,{y}^{4}{\hat{x}}^{4},
\eq
the remaining $19$ polynomials $\hat{p}_1$-$\hat{p}_{19}$ are at most quadratic in $\hat{x}$.
The polynomial $\hat{p}_{20}$ will lead to roots of a quartic polynomial as arguments of the multiple polylogarithms if the
differential equation is integrated along $\hat{x}$ for generic and constant $y$.

The master integrals for the sectors $B_{255}'$ and $B_{479}'$ are
\bq
 K_{1}
 & = &
 \eps^3  \; \left(1-w\right)\; {\bf D}^- I^{B'}_{101010100},
 \nonumber \\
 K_{2}
 & = &
 \eps^3 \; {\bf D}^- I^{B'}_{101\left(-1\right)10100},
 \nonumber \\
 K_{3}
 & = &
 \eps^3 \;r_2\; {\bf D}^- I^{B'}_{011010100},
 \nonumber \\
 K_{4}
 & = &
 \eps^3 \; \left[ {\bf D}^- I^{B'}_{\left(-1\right)11010100}+\left(1-w\right) {\bf D}^- I^{B'}_{011010100}\right],
 \nonumber \\
 K_{5}
 & = &
 \eps^3 \; \left[ {\bf D}^- I^{B'}_{01101\left(-1\right)100} +v \; {\bf D}^- I^{B'}_{011010100} \right],
 \nonumber \\
 K_{6}
 & = &
 \eps^3 \; \left[ {\bf D}^- I^{B'}_{0110101\left(-1\right)0} +v \; {\bf D}^- I^{B'}_{011010100} \right],
 \nonumber \\
 K_{7}
 & = &\eps^3\;v\;{\bf D}^- I^{B'}_{100011100},
 \nonumber \\
 K_{8}
 & = &
 \eps^3\;v\;{\bf D}^- I^{B'}_{001011100},
 \nonumber \\
 K_{9}
 & = &
 \eps^3\;v\;{\bf D}^- I^{B'}_{101000110},
 \nonumber \\
 K_{10}
 & = &
 \eps^3 \; r_3\;\left[ {\bf D}^- I^{B'}_{011010100}+{\bf D}^- I^{B'}_{111\left(-1\right)10100} + \left(1-w\right) \; {\bf D}^- I^{B'}_{111010100} \right],
 \nonumber \\
 K_{11}
 & = &
 \eps^3 \; r_3\;{\bf D}^- I^{B'}_{111\left(-1\right)10100} ,
 \nonumber \\
 K_{12}
 & = &
\eps^3 \;v\; r_3\;{\bf D}^- I^{B'}_{110011100} ,
 \nonumber \\
 K_{13}
 & = &
 \eps^3 \;r_2\;{\bf D}^- I^{B'}_{101\left(-1\right)11100},
 \nonumber \\
 K_{14}
 & = &
 \eps^3 \; \left(1-w\right) \left[ {\bf D}^- I^{B'}_{101\left(-1\right)11100} + v \; {\bf D}^- I^{B'}_{101011100} \right],
 \nonumber \\
 K_{15}
 & = &
 \eps^3 \;\left[ {\bf D}^- I^{B'}_{101\left(-2\right)11100} + v \; {\bf D}^- I^{B'}_{101\left(-1\right)11100} \right],
 \nonumber \\
 K_{16}
 & = &
 \eps^3 \; \left(1-w\right) \left[ v \; {\bf D}^- I^{B'}_{011011100} + {\bf D}^- I^{B'}_{011010100} \right],
 \nonumber \\
 K_{17}
 & = &
 \eps^3 \; v\;r_3 \; {\bf D}^- I^{B'}_{111000110},
 \nonumber \\
 K_{18}
 & = &
 \eps^3 \; v \;\left(1-w\right) \; {\bf D}^- I^{B'}_{101100110} ,
 \nonumber \\
 K_{19}
 & = &
 \eps^3 \; v \; r_2 \; {\bf D}^- I^{B'}_{011100110},
 \nonumber \\
 K_{20}
 & = &
 \eps^3 \; v \left[ {\bf D}^- I^{B'}_{\left(-1\right)11100110} + \left(1-w\right) \; {\bf D}^- I^{B'}_{011100110} \right],
 \nonumber \\
 K_{21}
 & = &
 \eps^3 \; v \; {\bf D}^- I^{B'}_{0111\left(-1\right)0110},
 \nonumber \\
 K_{22}
 & = &
 \eps^3 \; v^2 \; {\bf D}^- I^{B'}_{100101110},
 \nonumber \\
 K_{23}
 & = &
 \eps^3 \; v^2 \; {\bf D}^- I^{B'}_{001101110},
 \nonumber \\
 K_{24}
 & = &
 \eps^4 \left(1-2 \eps \right)\;v\; I^{B'}_{ 111021100},
 \nonumber \\
 K_{25}
 & = &
 \eps^3 \; r_3\;\left[\left(1-w\right){\bf D}^- I^{B'}_{111\left(-1\right)11100}+{\bf D}^- I^{B'}_{101\left(-1\right)11100} \right],
 \nonumber \\
 K_{26}
 & = &
 \eps^3 \; v\;r_3\;\left[\left(1-w\right){\bf D}^- I^{B'}_{111100110}+{\bf D}^- I^{B'}_{011100110} \right],
 \nonumber \\
 K_{27}
 & = &
 \eps^3 \; v\;r_3\;{\bf D}^- I^{B'}_{11101011\left(-1\right)},
 \nonumber \\
 K_{28}
 & = &
 \eps^4 \left(1-2 \eps \right)\;v\; I^{B'}_{112010110},
 \nonumber \\
 K_{29}
 & = &
 \eps^4 \left(1-2 \eps \right)\;v\;I^{B'}_{111020110},
 \nonumber \\
 K_{30}
 & = &
 \eps^3 \;v^2\;r_3\;{\bf D}^- I^{B'}_{110101110},
 \nonumber \\
 K_{31}
 & = &
 \eps^5 \left(1-2 \eps \right)\;v \; I^{B'}_{111110110},
 \nonumber \\
  K_{32}
 & = &
\eps^4 \left(1-2 \eps \right)^2\;v \;  I^{B'}_{111101110},
 \nonumber \\
 K_{33}
 & = &
 \eps^3  \; {\bf D}^- I^{B'}_{101000001},
 \nonumber \\
 K_{34}
 & = &
 \eps^3 \; r_3 \; {\bf D}^- I^{B'}_{111000001},
 \nonumber \\
 K_{35}
 & = &
 \eps^3 \;\left(1-w\right) {\bf D}^- I^{B'}_{101100001},
 \nonumber \\
 K_{36}
 & = &
 \eps^3 \;r_2\; {\bf D}^- I^{B'}_{011100001},
 \nonumber \\
 K_{37}
 & = &
 \eps^3 \left[{\bf D}^- I^{B'}_{\left(-1\right)11100001}+\left(1-w\right){\bf D}^- I^{B'}_{011100001}\right] ,
 \nonumber \\
 K_{38}
 & = &
 \eps^3 \;  \left[ {\bf D}^- I^{B'}_{01110\left(-1\right)001}
   + v\; {\bf D}^- I^{B'}_{011100001}
 \right],
 \nonumber \\
 K_{39}
 & = &
 \eps^2 \; \left(1+4 \eps\right) \; {\bf D}^- I^{B'}_{001100101},
 \nonumber \\
 K_{40}
 & = &
 \eps^3 \; r_1 \; {\bf D}^- I^{B'}_{001100011},
 \nonumber \\
 K_{41}
 & = &
 \eps^3 \; \left[ {\bf D}^- I^{B'}_{\left(-1\right)01100011} +\left(1-w\right)  \; {\bf D}^- I^{B'}_{001100011} \right],
 \nonumber \\
 K_{42}
 & = &
 \eps^3 \; \left[ {\bf D}^- I^{B'}_{0011\left(-1\right)0011}
  + v \; {\bf D}^- I^{B'}_{001100011}
 \right],
 \nonumber \\
 K_{43}
 & = &
 \eps^3 \; v \; {\bf D}^-I^{B'}_{001000111},
 \nonumber \\
 K_{44}
 & = &
 \eps^3 \; r_3 \left[ \left(1-w\right) \; {\bf D}^- I^{B'}_{111100001} + {\bf D}^- I^{B'}_{011100001} \right],
 \nonumber \\
 K_{45}
 & = &
 \eps^3 \; \left(1-w\right) 
  \left[ \left(1-w\right) \; {\bf D}^- I^{B'}_{101100101} + {\bf D}^- I^{B'}_{001100101} \right],
 \nonumber \\
 K_{46}
 & = &
 \eps^3 \; r_2 \left[ {\bf D}^- I^{B'}_{(-1)11100101} + \left(1-w\right) \; {\bf D}^- I^{B'}_{011100101} \right],
 \nonumber \\
 K_{47}
 & = &
 \eps^3 \left[ {\bf D}^- I^{B'}_{(-2)11100101} + 2 \left(1-w\right) \; {\bf D}^- I^{B'}_{(-1)11100101} + \left(1-w\right)^2 \; {\bf D}^- I^{B'}_{011100101} \right],
 \nonumber \\
 K_{48}
 & = &
 \eps^3 \; \left(w-v\right) \left[ {\bf D}^- I^{B'}_{0111(-1)0101} - 2 \; {\bf D}^- I^{B'}_{011000101} \right],
 \nonumber \\
 K_{49}
 & = &
 \eps^3 \; r_2 \left[ \left(1-w\right) \; {\bf D}^- I^{B'}_{101100011} + {\bf D}^- I^{B'}_{001100011} \right],
 \nonumber \\
 K_{50}
 & = &
 \eps^3 \; \left(1-w\right) \left[ {\bf D}^- I^{B'}_{1(-1)1100011} + \left(1-w\right) \; {\bf D}^- I^{B'}_{101100011} \right],
 \nonumber \\
 K_{51}
 & = &
 \eps^3 \; \left(1-w\right) \left[ {\bf D}^- I^{B'}_{101100(-1)11} + v \; {\bf D}^- I^{B'}_{101100011} \right],
 \nonumber \\
 K_{52}
 & = &
 \eps^3 \; r_3 \left[ \left(1-w\right)^2 \; {\bf D}^- I^{B'}_{111100101} + 2 \left(1-w\right) \; {\bf D}^- I^{B'}_{011100101} + {\bf D}^- I^{B'}_{(-1)11100101} \right],
 \nonumber \\
 K_{53}
 & = &
 \eps^3 \; r_3 \left[ \left(1-w\right)^2 \; {\bf D}^- I^{B'}_{111100011} + \left(1-w\right) \left( {\bf D}^- I^{B'}_{011100011} + {\bf D}^- I^{B'}_{101100011} \right) + {\bf D}^- I^{B'}_{001100011} \right],
 \nonumber \\
 K_{54}
 & = &
 \eps^4 \left(1-2\eps\right) \; v \; I^{B'}_{101110021},
 \nonumber \\
 K_{55}
 & = &
 \eps^4 \left(1-2\eps\right) \; v \; I^{B'}_{112000111},
 \nonumber \\
 K_{56}
 & = &
 \eps^4 \left(1-2\eps\right) \; v \; I^{B'}_{012100111},
 \nonumber \\
 K_{57}
 & = &
 \eps^4 \left(1-2\eps\right) \; v \; I^{B'}_{011200111},
 \nonumber \\
 K_{58}
 & = &
 \eps^5 \left(1-2\eps\right) \; v \; I^{B'}_{111110011},
 \nonumber \\
 K_{59}
 & = &
 \eps^4 \left(1-2\eps\right) \; v  \; I^{B'}_{112110011},
 \nonumber \\
 K_{60}
 & = &
 \eps^4 \left(1-2\eps\right) \; r_3 \left[ w\;I^{B'}_{211110011} + I^{B'}_{112110011} - 2 \eps \; I^{B'}_{111110011} \right],
 \nonumber \\
 K_{61}
 & = &
 \eps^4 \left(1-2\eps\right) \; v w \; I^{B'}_{112100111}.
\eq
As boundary point we take again the point $(v,w)=(0,1)$.
We denote the boundary values for this system by $C'$.
The non-zero boundary values are
\bq
 & &
 C_{33}'
 \; = \;
 C_{38}'
 \; = \; C_{1},
 \nonumber \\
 & &
 C_{9}'
 \; = \;
 C_{21}'
 \; = \;
 C_{43}'
 \; = \; C_{15},
 \nonumber \\
 & &
 C_{7}'
 \; = \;
 C_{8}'
 \; = \; C_{51},
 \nonumber \\
 & &
 C_{22}'
 \; = \;
 C_{23}'
 \; = \; C_{60},
 \nonumber \\
 & &
 C_{39}'
 \; = \; C_{71},
 \nonumber \\
 & &
 C_{41}'
 \; = \; 
 C_{47}'
 \; = \; \frac{1}{3} C_{71},
 \nonumber \\
 & &
 C_{2}'
 \; = \;
 C_{5}'
 \; = \;
 C_{15}'
 \; = \;
 C_{42}'
 \; = \; -\frac{2}{3} C_{71},
 \nonumber \\
 & &
 C_{48}'
 \; = \; \frac{1}{2} C_{1} - \frac{1}{2} C_{71}.
\eq

\subsection{The sector $B_{509}'$}

The sector $B_{509}'$ involves the square roots $(r_1,r_2,r_4)$. 
These are rationalised by the variables $(\tilde{x},y)$.
The variable $\tilde{x}$ is defined by
\bq
 x \; = \;\tilde{x} \frac{\left(1-\tilde{x}\right)}{\left(1+\tilde{x}\right)},
 & &
 \tilde{x}
 \; = \; 
 \frac{1}{2} \left( 1-x-\sqrt{x^2-6x+1}\right).
\eq
The variables $x$ and $y$ are defined in eq.~(\ref{variable_trafo_inv}).
The variable $\tilde{x}$ rationalises simultaneously $r_1$ and $r_3$ and has already appeared in \cite{Adams:2018kez}.
The root $r_4$ is given in the variable $\tilde{x}$ by
\bq
 r_4
 & = &
 \frac{\left(1+\tilde{x}^2\right)\left(1-2\tilde{x}-\tilde{x}^2\right)}{\tilde{x}\left(1-\tilde{x}^2\right)}.
\eq
The value $x=1+i\delta$ (with $\delta$ being an infinitesimal small positive number) corresponds to $\tilde{x}=i$.
We set
\bq
 \tilde{x}' & = & i - \tilde{x}.
\eq
With this definition the value $x=1+i\delta$ corresponds to $\tilde{x}'=0$.

With this rationalisation the differential one-forms translate to
\bq
 d\ln\left(\tilde{p}_{1}\right), \dots, d\ln\left(\tilde{p}_{17}\right),
\eq
where $\tilde{p}_{1}$-$\tilde{p}_{17}$ are polynomials in $(\tilde{x},y)$.
The differential one-form $\omega_6=d\ln(w-v)$ leads also with this rationalisation to a quartic polynomial in $\tilde{x}$:
\bq
 \tilde{p}_{17}
 =
 1-2\,y+\tilde{x}+{y}^{2}-4\,y\tilde{x}+2\,{\tilde{x}}^{2}-{y}^{2}\tilde{x}-{\tilde{x}}^{3}+2\,{y}^{2}{\tilde{x}}^{2}-4\,{\tilde{x}}^{3}y+{\tilde{x}}^{4}+{\tilde{x}}^{3}{y}^{2}+2\,{\tilde{x}}^{4}y+{\tilde{x}}^{4}{y}^{2},
\eq
the remaining $16$ polynomials $\tilde{p}_1$-$\tilde{p}_{16}$ are at most quadratic in $\tilde{x}$.
We note that $\tilde{p}_{17}$ (and all other polynomials $\tilde{p}_1$-$\tilde{p}_{16}$) are quadratic in $y$.

The master integrals for the sector $B_{509}'$ are
\bq
 L_{1}
 & = &
 K_{1} \; = \;
 \eps^3  \; \left(1-w\right)\; {\bf D}^- I^{B'}_{101010100},
 \nonumber \\
 L_{2}
 & = &
 K_{2} \; = \;
 \eps^3 \; {\bf D}^- I^{B'}_{101\left(-1\right)10100},
 \nonumber \\
 L_{3}
 & = &
 K_{3} \; = \;
 \eps^3 \;r_2\; {\bf D}^- I^{B'}_{011010100},
 \nonumber \\
 L_{4}
 & = &
 K_{4} \; = \;
 \eps^3 \; \left[ {\bf D}^- I^{B'}_{\left(-1\right)11010100}+\left(1-w\right) {\bf D}^- I^{B'}_{011010100}\right],
 \nonumber \\
 L_{5}
 & = &
 K_{5} \; = \;
 \eps^3 \; \left[ {\bf D}^- I^{B'}_{01101\left(-1\right)100} +v \; {\bf D}^- I^{B'}_{011010100} \right],
 \nonumber \\
 L_{6}
 & = &
 K_{6} \; = \;
 \eps^3 \; \left[ {\bf D}^- I^{B'}_{0110101\left(-1\right)0} +v \; {\bf D}^- I^{B'}_{011010100} \right],
 \nonumber \\
 L_{7}
 & = &
 K_{7} \; = \;
 \eps^3\;v\;{\bf D}^- I^{B'}_{100011100},
 \nonumber \\
 L_{8}
 & = &
 K_{8} \; = \;
 \eps^3\;v\;{\bf D}^- I^{B'}_{001011100},
 \nonumber \\
 L_{9}
 & = &
 K_{9} \; = \;
 \eps^3\;v\;{\bf D}^- I^{B'}_{101000110},
 \nonumber \\
 L_{10}
 & = &
 K_{13} \; = \;
 \eps^3 \;r_2\;{\bf D}^- I^{B'}_{101\left(-1\right)11100},
 \nonumber \\
 L_{11}
 & = &
 K_{14} \; = \;
 \eps^3 \; \left(1-w\right) \left[ {\bf D}^- I^{B'}_{101\left(-1\right)11100} + v \; {\bf D}^- I^{B'}_{101011100} \right],
 \nonumber \\
 L_{12}
 & = &
 K_{15} \; = \;
 \eps^3 \;\left[ {\bf D}^- I^{B'}_{101\left(-2\right)11100} + v \; {\bf D}^- I^{B'}_{101\left(-1\right)11100} \right],
 \nonumber \\
 L_{13}
 & = &
 K_{16} \; = \;
 \eps^3 \; \left(1-w\right) \left[ v \; {\bf D}^- I^{B'}_{011011100} + {\bf D}^- I^{B'}_{011010100} \right],
 \nonumber \\
 L_{14}
 & = &
 K_{18} \; = \;
 \eps^3 \; v \;\left(1-w\right) \; {\bf D}^- I^{B'}_{101100110} ,
 \nonumber \\
 L_{15}
 & = &
 K_{19} \; = \;
 \eps^3 \; v \; r_2 \; {\bf D}^- I^{B'}_{011100110},
 \nonumber \\
 L_{16}
 & = &
 K_{20} \; = \;
 \eps^3 \; v \left[ {\bf D}^- I^{B'}_{\left(-1\right)11100110} + \left(1-w\right) \; {\bf D}^- I^{B'}_{011100110} \right],
 \nonumber \\
 L_{17}
 & = &
 K_{21} \; = \;
 \eps^3 \; v \; {\bf D}^- I^{B'}_{0111\left(-1\right)0110},
 \nonumber \\
 L_{18}
 & = &
 K_{22} \; = \;
 \eps^3 \; v^2 \; {\bf D}^- I^{B'}_{100101110},
 \nonumber \\
 L_{19}
 & = &
 K_{23} \; = \;
 \eps^3 \; v^2 \; {\bf D}^- I^{B'}_{001101110},
 \nonumber \\
 L_{20}
 & = &
 K_{33} \; = \;
 \eps^3  \; {\bf D}^- I^{B'}_{101000001},
 \nonumber \\
 L_{21}
 & = &
 K_{35} \; = \;
 \eps^3 \;\left(1-w\right) {\bf D}^- I^{B'}_{101100001},
 \nonumber \\
 L_{22}
 & = &
 K_{36} \; = \;
 \eps^3 \;r_2\; {\bf D}^- I^{B'}_{011100001},
 \nonumber \\
 L_{23}
 & = &
 K_{37} \; = \;
 \eps^3 \left[{\bf D}^- I^{B'}_{\left(-1\right)11100001}+\left(1-w\right){\bf D}^- I^{B'}_{011100001}\right] ,
 \nonumber \\
 L_{24}
 & = &
 K_{38} \; = \;
 \eps^3 \;  \left[ {\bf D}^- I^{B'}_{01110\left(-1\right)001}
   + v\; {\bf D}^- I^{B'}_{011100001}
 \right],
 \nonumber \\
 L_{25}
 & = &
 K_{39} \; = \;
 \eps^2 \; \left(1+4 \eps\right) \; {\bf D}^- I^{B'}_{001100101},
 \nonumber \\
 L_{26}
 & = &
 K_{40} \; = \;
 \eps^3 \; r_1 \; {\bf D}^- I^{B'}_{001100011},
 \nonumber \\
 L_{27}
 & = &
 K_{41} \; = \;
 \eps^3 \; \left[ {\bf D}^- I^{B'}_{\left(-1\right)01100011} +\left(1-w\right)  \; {\bf D}^- I^{B'}_{001100011} \right],
 \nonumber \\
 L_{28}
 & = &
 K_{42} \; = \;
 \eps^3 \; \left[ {\bf D}^- I^{B'}_{0011\left(-1\right)0011}
  + v \; {\bf D}^- I^{B'}_{001100011}
 \right],
 \nonumber \\
 L_{29}
 & = &
 K_{43} \; = \;
 \eps^3 \; v \; {\bf D}^-I^{B'}_{001000111},
 \nonumber \\
 L_{30}
 & = &
 K_{45} \; = \;
 \eps^3 \; \left(1-w\right) 
  \left[ \left(1-w\right) \; {\bf D}^- I^{B'}_{101100101} + {\bf D}^- I^{B'}_{001100101} \right],
 \nonumber \\
 L_{31}
 & = &
 K_{46} \; = \;
 \eps^3 \; r_2 \left[ {\bf D}^- I^{B'}_{(-1)11100101} + \left(1-w\right) \; {\bf D}^- I^{B'}_{011100101} \right],
 \nonumber \\
 L_{32}
 & = &
 K_{47} \; = \;
 \eps^3 \left[ {\bf D}^- I^{B'}_{(-2)11100101} + 2 \left(1-w\right) \; {\bf D}^- I^{B'}_{(-1)11100101} + \left(1-w\right)^2 \; {\bf D}^- I^{B'}_{011100101} \right],
 \nonumber \\
 L_{33}
 & = &
 K_{48} \; = \;
 \eps^3 \; \left(w-v\right) \left[ {\bf D}^- I^{B'}_{0111(-1)0101} - 2 \; {\bf D}^- I^{B'}_{011000101} \right],
 \nonumber \\
 L_{34}
 & = &
 K_{49} \; = \;
 \eps^3 \; r_2 \left[ \left(1-w\right) \; {\bf D}^- I^{B'}_{101100011} + {\bf D}^- I^{B'}_{001100011} \right],
 \nonumber \\
 L_{35}
 & = &
 K_{50} \; = \;
 \eps^3 \; \left(1-w\right) \left[ {\bf D}^- I^{B'}_{1(-1)1100011} + \left(1-w\right) \; {\bf D}^- I^{B'}_{101100011} \right],
 \nonumber \\
 L_{36}
 & = &
 K_{51} \; = \;
 \eps^3 \; \left(1-w\right) \left[ {\bf D}^- I^{B'}_{101100(-1)11} + v \; {\bf D}^- I^{B'}_{101100011} \right],
 \nonumber \\
 L_{37}
 & = &
 K_{54} \; = \;
 \eps^4 \left(1-2\eps\right) \; v \; I^{B'}_{101110021},
 \nonumber \\
 L_{38}
 & = &
 K_{56} \; = \;
 \eps^4 \left(1-2\eps\right) \; v \; I^{B'}_{012100111},
 \nonumber \\
 L_{39}
 & = &
 K_{57} \; = \;
 \eps^4 \left(1-2\eps\right) \; v \; I^{B'}_{011200111},
 \nonumber \\
 L_{40}
 & = &
 \eps^3 \; r_4 \left[ v \; {\bf D}^- I^{B'}_{0011(-1)1111} + 2 \; {\bf D}^- I^{B'}_{001100011} \right],
 \nonumber \\
 L_{41}
 & = &
 \eps^4 \left(1-2\eps\right) \; v \; I^{B'}_{002101111},
 \nonumber \\
 L_{42}
 & = &
 \eps^5 \left(1-2\eps\right) \; v \; I^{B'}_{101101111},
 \nonumber \\
 L_{43}
 & = &
 \eps^4 \left(1-2\eps\right) \; v \left(1-w\right) \; I^{B'}_{102101111}.
\eq
As boundary point we take again the point $(v,w)=(0,1)$.
We denote the boundary values for this system by $C''$.
The non-zero boundary values are
\bq
 & &
 C_{20}''
 \; = \;
 C_{24}''
 \; = \; C_{1},
 \nonumber \\
 & &
 C_{9}''
 \; = \;
 C_{17}''
 \; = \;
 C_{29}''
 \; = \; C_{15},
 \nonumber \\
 & &
 C_{7}''
 \; = \;
 C_{8}''
 \; = \; C_{51},
 \nonumber \\
 & &
 C_{18}''
 \; = \;
 C_{19}''
 \; = \; C_{60},
 \nonumber \\
 & &
 C_{25}''
 \; = \; C_{71},
 \nonumber \\
 & &
 C_{27}''
 \; = \;
 C_{32}''
 \; = \; \frac{1}{3} C_{71},
 \nonumber \\
 & &
 C_{2}''
 \; = \; 
 C_{5}''
 \; = \; 
 C_{12}''
 \; = \; 
 C_{28}''
 \; = \; -\frac{2}{3} C_{71},
 \nonumber \\
 & &
 C_{33}''
 \; = \; \frac{1}{2} C_{1} - \frac{1}{2} C_{71}.
\eq

\section{Supplementary material}
\label{sect:supplement}

Attached to the arxiv version of this article is an electronic file
in ASCII format with {\tt Maple} syntax, defining the quantities
\begin{center}
 \verb|A|,
 \;
 \verb|B|,
 \;
 \verb|C|,
 \;
 \verb|J|.
\end{center}
The matrix \verb|A| appears
in the differential equation
\bq
 d \vec{J}
 & = &
 A \vec{J}.
\eq
The entries of the matrix $A$ are linear combinations of $\omega_1$, ..., $\omega_{19}$, defined in eq.~(\ref{def_differential_one_forms}).
These differential forms are denoted
by
\begin{center}
 \verb|omega_1|, ..., \verb|omega_19|.
\end{center}
The matrices $B$ and $C$ appear in the differential equations in eq.~(\ref{diff_eq_K_L}).

The vector \verb|J| contains the results for the master integrals up to order $\eps^5$ in terms of multiple polylogarithms.
The variable $\eps$ is denoted by \verb|eps|, $\zeta_2$, $\zeta_3$, $\zeta_5$ by
\begin{center}
 \verb|zeta_2|, \verb|zeta_3|, \verb|zeta_5|,
\end{center}
respectively.
For the notation of multiple polylogarithms we give an example: $G(x_7',x_8',1;x')$ is denoted by
\begin{center}
  \verb|Glog([xp7,xp8,1],xp)|.
\end{center}
The file size is $10 \, \mathrm{MB}$, roughly $90 \%$ of the file size is used 
for the $\eps^5$-term of $J$.
The file size increases approximately by a factor of $10$ for every additional order.
This is expected: With an alphabet of
\bq
 \Nletter & = & 21
\eq
letters we have
\bq
 \left(\Nletter\right)^w
\eq
possible multiple polylogarithms at weight $w$.
As the matrix $A$ is sparse, the actual increase factor is of the order ${\mathcal O}(10)$ 
and roughly a factor of two smaller than $\Nletter=21$.
Note that the explicit expressions for $J$ are convenient, but not essential:
All information to any order in $\eps$ is stored compactly in the matrix $A$
and the boundary values given in section~\ref{sect:boundary}.

\end{appendix}

{\footnotesize
\bibliography{/home/stefanw/notes/biblio}
\bibliographystyle{/home/stefanw/latex-style/h-physrev5}
}

\end{document}